 \documentclass[twocolumn,twocolappendix]{aastex631}

\shorttitle{High-resolution far-IR survey}
\shortauthors{Bonato et al.}
\graphicspath{{./}{figures/}}
\begin{document}

\title{A High-resolution Far-infrared Survey to Probe Black Hole-Galaxy Co-evolution}

\author[0000-0001-9139-2342]{Matteo Bonato}
\affiliation{INAF--Istituto di Radioastronomia and Italian ALMA Regional Centre, Via Gobetti 101, Bologna, Italy, I-40129}
\email{matteo.bonato@inaf.it}

\author{David Leisawitz}
\affiliation{Goddard Space Flight Center, 8800 Greenbelt Rd., Greenbelt, MD USA 20771}

\author{Gianfranco De Zotti}
\affiliation{INAF - Osservatorio Astronomico di Padova, Vicolo dell’Osservatorio 5, I-35122 Padova, Italy}

\author{Laura Sommovigo}
\affiliation{Center for Computational Astrophysics, Flatiron Institute, 162 5th Avenue,
New York, NY 10010, USA}

\author{Irene Shivaei}
\affiliation{Centro de Astrobiología (CAB), CSIC-INTA, Ctra. de Ajalvir km 4, Torrejón de Ardoz, E-28850,
Madrid, Spain}

\author{C. Megan Urry}
\affiliation{Yale University, P.O. Box 208120, New Haven, CT USA 06520-8120}

\author{Duncan Farrah}
\affiliation{University of Hawai’i at Mānoa, Honolulu County, Hawaii, USA 96822}

\author{Locke Spencer}
\affiliation{University of Lethbridge, Lethbridge, Alberta T1K 3M4, Canada}

\author{Berke V. Ricketti}
\affiliation{University of Lethbridge, Lethbridge, Alberta T1K 3M4, Canada}

\author{Hannah Rana}
\affiliation{Center for Astrophysics $|$ Harvard \& Smithsonian, 60 Garden Street, Cambridge, MA 02138, USA}

\author{Susanne Aalto}
\affiliation{Department of Space, Earth and Environment, Chalmers University of Technology, 412 96 Göteborg, Sweden}

\author[0000-0002-1233-9998]{David B. Sanders}
\affiliation{University of Hawai’i at Mānoa, Honolulu County, Hawaii, USA 96822}

\author{Lee G. Mundy}
\affiliation{University of Maryland, College Park, MD USA 20742}

%% Note that the \and command from previous versions of AASTeX is now
%% depreciated in this version as it is no longer necessary. AASTeX
%% automatically takes care of all commas and "and"s between authors names.

%% AASTeX 6.31 has the new \collaboration and \nocollaboration commands to
%% provide the collaboration status of a group of authors. These commands
%% can be used either before or after the list of corresponding authors. The
%% argument for \collaboration is the collaboration identifier. Authors are
%% encouraged to surround collaboration identifiers with ()s. The
%% \nocollaboration command takes no argument and exists to indicate that
%% the nearby authors are not part of surrounding collaborations.

%% Mark off the abstract in the ``abstract'' environment.
\begin{abstract}
Far-infrared (FIR) surveys are critical to probing the co-evolution of black holes and galaxies, since of order half the light from accreting black holes and active star formation is emitted in the rest-frame infrared over $0.5\lesssim z \lesssim 10$.
For deep fields with areas of 1 deg$^2$ or less, like the legacy surveys GOODS, COSMOS, and CANDELS, source crowding means that sub-arcsecond resolution is essential. In this paper, we show with a simulation of the FIR sky that observations made with a small telescope (2 m) at low angular resolution preferentially detect the brightest galaxies, and we demonstrate the scientific value of a space mission that would offer sub-arcsecond resolution. We envisage a facility that would provide high-resolution imaging and spectroscopy over the wavelength range $25 - 400\,\mu$m, and we present predictions for an extragalactic survey covering $0.5\,\hbox{deg}^2$. Such a survey is expected to detect tens of thousands of star-forming galaxies and thousands of Active Galactic Nuclei (AGN), in multiple FIR lines (e.g.  [CII], [OI], [CI]) and continuum.  At the longest wavelengths (200--400$\,\mu$m), it would probe beyond the reionization epoch, up to $z\sim 7$--8. A combination of spectral resolution, line sensitivity, and broad spectral coverage would allow us to learn about the physical conditions (temperature, density, metallicity) characterizing the interstellar medium of galaxies over the past $\sim 12$ billion years and to investigate galaxy-AGN co-evolution.

\end{abstract}

%% Keywords should appear after the \end{abstract} command.
%% The AAS Journals now uses Unified Astronomy Thesaurus concepts:
%% https://astrothesaurus.org
%% You will be asked to selected these concepts during the submission process
%% but this old "keyword" functionality is maintained in case authors want
%% to include these concepts in their preprints.

%% From the front matter, we move on to the body of the paper.
%% Sections are demarcated by \section and \subsection, respectively.
%% Observe the use of the LaTeX \label
%% command after the \subsection to give a symbolic KEY to the
%% subsection for cross-referencing in a \ref command.
%% You can use LaTeX's \ref and \label commands to keep track of
%% cross-references to sections, equations, tables, and figures.
%% That way, if you change the order of any elements, LaTeX will
%% automatically renumber them.
%%
%% We recommend that authors also use the natbib \citep
%% and \citet commands to identify citations.  The citations are
%% tied to the reference list via symbolic KEYs. The KEY corresponds
%% to the KEY in the \bibitem in the reference list below.

%\begin{spacing}{2}   % use double spacing for rest of manuscript

\section{Introduction}\label{sect:introduction}

Multiwavelength surveys have provided great insight into the evolution of galaxies and Active Galactic Nuclei (AGN) out to the Epoch of Reionization (EoR; z$\sim$6-7; see \citealt{Venemans2020}, \citealt{Bouwens2020}; see also \citealt{Hodge_daCunha2020} for a review on high-z sub-mm observations and e.g. \citealt{Finkelstein2023} and \citealt{Eisenstein2023} for recent surveys with the James Webb Space Telescope (JWST)). These objects exhibit a broad range of rest-frame spectral energy distributions (SEDs), from far-infrared (FIR) to ultraviolet energies for galaxies and from radio through gamma-ray energies for AGN. Thus it is vital to have broad wavelength coverage in a wide redshift range in order to reliably quantify the total energy emitted by these sources and the physical mechanisms involved. The FIR (defined here as 25 - 400\,$\mu$m) is particularly important because dust obscures most star-forming galaxies \citep{Lagache2005, Lemaux2014, PerezTorres2021} and AGN \citep{treister2004, TruebenbachDarling2017, HickoxAlexander2018, ananna2019, Laloux2023}, so even though they generate most of their primary radiation in the ultraviolet, we observe reprocessed light emitted primarily at FIR wavelengths.
That is, most of the detectable energy emitted from stars and black hole accretion comes out in the FIR at $z\leq 3$ (see \citealt{MadauDickinson2014}). {Cosmic star formation and black hole (BH) accretion histories are dominated by the FIR emission coming from dusty star forming galaxies (SFGs) and AGN at $z\leq 3$.} Without this crucial waveband, we cannot understand the energetics of AGN, the processes of star formation, or the co-evolution of black holes and galaxies  \citep{Lutz2014, Pouliasis2020, Thorne2022, LyuRieke2022, Auge2022, Laloux2023}.

Multi-wavelength surveys from the past two decades span an extensive range in luminosity and redshift, probing massive sources ($M_{\star}>10^9\ \mathrm{M_{\odot}}$) out to $z\sim 4-6$ (galaxies: e.g. \citealt{LeFevre2020,Faisst2020,Bouwens2022,Carnall2023}, AGN: e.g. \citealt{Venemans2020,Kocevski2023,Scholtz2023}).
%A ``wedding cake'' strategy can help to explore the full population of black holes and galaxies, combining wide and shallow surveys with those sampling narrow fields deeply. This is because any flux-limited survey produces a sample in which luminosity and redshift are strongly correlated. Wide-area surveys at shallow depths detect rare sources like massive galaxies or luminous AGN, while smaller, deeper surveys find lower luminosity AGN and galaxies with luminosities at or below $L_\star$. The combination of a wide range of volumes provides us with a better understanding of the full population of galaxies and AGN.

{Wide}, shallow surveys in the mid- and far-infrared have been done with the Wide-Field Infrared Survey Explorer \citep[WISE;][]{Wright2010} and the \textit{Herschel} Spectral and Photometric Imaging REceiver \citep[SPIRE;][]{Eales2010, Oliver2012, Viero2014}. But the deep fields,
like the Great Observatories Origins Deep Survey \citep[GOODS, centered on the Chandra Deep Fields South and North;][]{Giavalisco2004, treister2004,treister2006a,xue2012,lehmer2012,ranalli2013,luo2017,liu2017}, the Cosmic Evolution Survey  \citep[COSMOS;][]{scoville2007, sanders2007, hasinger2007, cappelluti2007, cappelluti2009, elvis2009, civano2012, civano2016, Smolcic2017a, Smolcic2017b, Weaver2022, Heywood2022}, the Cosmic Assembly Near-infrared Deep Extragalactic Legacy Survey \citep[CANDELS;][]{Grogin2011, Koekemoer2011, Galametz2013, Guo2013, Stefanon2017, Nayyeri2017, Barro2019}, the All-wavelength Extended Groth strip International Survey \citep[AEGIS;][]{Davis2007, Konidaris2007, Symeonidis2007, Laird2009, Nandra2015}, the surveys of the Lockman Hole  and ELAIS regions \citep{Hasinger2001, Werner2004, Polletta2006, Coppin2006, GonzalezSolares2011, Mauduit2012, Oliver2012, Prandoni2018, Kondapally2021, Bonato2021}, reach flux levels where FIR-emitting sources are heavily confused at resolutions greater than a few arcseconds.

Even a large single-dish infrared telescope, like the 3.5-meter \textit{Herschel} Space telescope, has a spatial resolution in the tens of arcseconds, hopeless for resolving individual sources in the deeper fields. Moreover, precise positions are needed to match FIR sources with their counterparts at optical and X-ray energies, as the full spectral energy distribution (SED) is needed to disentangle emission from starlight and from accretion.
%For decades, sub-arcsecond resolution has been available in the optical\footnote{\textbf{e.g. in the Sloan Digital Sky Survey (SDSS; \citealt{York2000}) and in the Physics at High Angular Resolution in Nearby Galaxies Survey (PHANGS; \citealt{Lee2022})}} and X-ray\footnote{\textbf{e.g. in the surveys conducted with the Chandra X-ray Observatory (CXO; \citealt{Weisskopf2002}) and with the Spectrum-Roentgen-Gamma (SRG)/ART-XC (\citealt{Pavlinsky2021})}}, but not the FIR.
{For decades, sub-arcsecond resolution has been available for surveys in the optical, thanks to the Hubble Space Telescope \citep{Williams1996, Williams2000}, and X-ray, thanks to the \textit{Chandra X-Ray Observatory} \citep{Weisskopf2000, Brandt2001}, but not the FIR.}

\begin{figure}
\begin{center}
\begin{tabular}{c}
\includegraphics[trim=0.0cm 0.0cm 0.4cm 0.0cm,clip=true,height=5.5cm, angle=0]{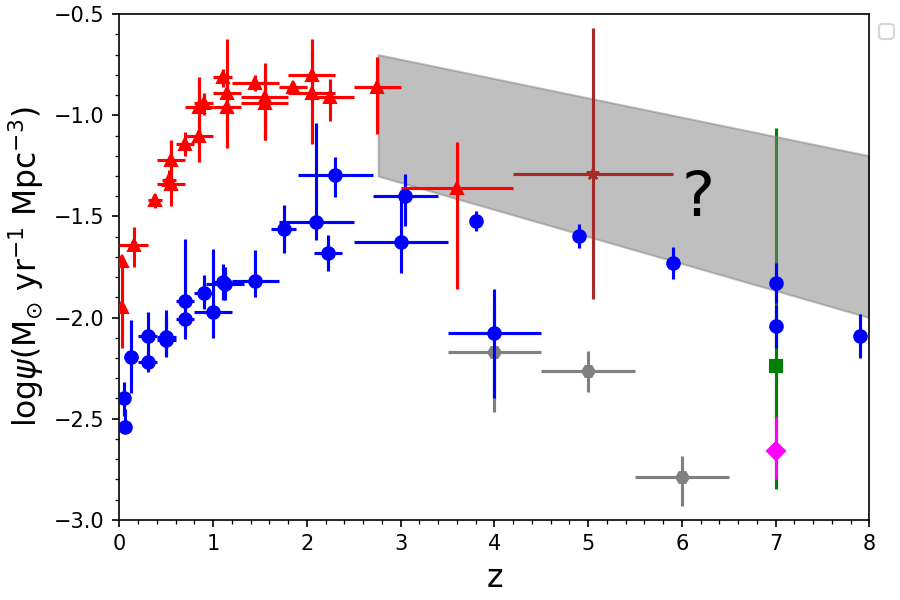}
\end{tabular}
\end{center}
\caption
{\label{fig:SFRH}
Evolution of the cosmic star-formation history from UV data (uncorrected for dust extinction) and from FIR data (blue circles and red triangles, respectively; data taken from the \citealt{MadauDickinson2014} compilation). Because of dust obscuration, UV surveys miss a large fraction of the cosmic star formation, recovered by FIR data. The latter however is largely missing at $z \gtrsim 3$ due to the strong confusion limits affecting \textit{Herschel}/SPIRE surveys. As a consequence, our understanding of the cosmic star formation history is incomplete. The reference survey described in this paper would extend measurements up to the reionization epoch. As further examples of more recent measurements, we also plot data points from: \citet{Casey2021} (the grey hexagons), derived from the ALMA ``MORA'' survey on a sample of 2-mm selected dusty SFGs; from \citet{Loiacono2021} (``field'' case; the brown star), derived from the ALMA ``ALPINE'' survey on a sample of serendipitous [CII] emitters; from \citet{Algera2023} ({results based on their Monte Carlo approach}; the green square) and \citet{Barrufet2023} (the magenta diamond) derived from the ALMA ``REBELS'' survey on an UV-selected galaxy sample.}
\end{figure}

Due to the limited sensitivity and the relatively bright confusion limits of \textit{Spitzer}\footnote{{The \textit{Spitzer}  InfraRed Array Camera (IRAC) $5\,\sigma$ point-source sensitivity for a 200\,s exposure time in low background is 2.0, 4.2, 27.5, and $34.5\,\mu$Jy for the  3.6, 4.5, 5.8 and $8.0\,\mu$m channels, respectively (IRAC Instrument Handbook; \url{https://irsa.ipac.caltech.edu/data/SPITZER/docs/irac/iracinstrumenthandbook/IRAC_Instrument_Handbook.pdf}).  The $5\,\sigma$ confusion limits for the Multiband Imaging Photometer for \textit{Spitzer} (MIPS) were estimated by \citet{Dole2004} to be $56\,\mu$Jy, 3.2\,mJy, and 40\,mJy at 24, 70, and $160\,\mu$m, respectively.}} and \textit{Herschel}\footnote{{The $5\,\sigma$ confusion limits of the \textit{Herschel} Photodetector Array Camera and Spectrometer (PACS)  are 0.4\,mJy in the $70\,\mu$m passband \citep{Berta2011},  0.75 and 3.4\,mJy at 100 and $160\,\mu$m, respectively \citep{Magnelli2013}. Those of SPIRE are 29, 31.5 and 34 mJy/beam at 250, 350 and $500\,\mu$m, respectively \citep{Nguyen2010}.}}, dust-obscured star formation in typical galaxies is well constrained only up to redshifts $z\sim 2.5-3$ \citep{Reddy2012, Gruppioni2013, MadauDickinson2014, Alvarez-Marquez2016, Shivaei2017}. At higher-redshift, the selection bias towards the most infrared-luminous and/or gravitationally lensed galaxies \citep{Negrello2010, Riechers2013, Negrello2014, PlanckCollaboration2015highz, Oteo2016, Negrello2017, Marrone2018, Greenslade2020} as well as the lack of multiple-FIR band detections of large galaxy samples, casts large uncertainties on Star Formation Rate Density (SFRD) studies (see e.g. \citealt{Gruppioni2020, Casey2021, Loiacono2021, Algera2023, Barrufet2023} for the latest constraints on the dust-obscured SFRD at $z=4-7$). Thus our view of cosmic star formation at early epochs is still far from complete (Fig.\,\ref{fig:SFRH}).

Recent JWST observations have significantly enhanced our understanding of the unobscured star formation rate in galaxies out to unprecedentedly early epochs (redshift $z > 10$, see e.g., \citealt{Harikane2023}; \citealt{Donnan2023a}; \citealt{Finkelstein2023}; \citealt{Casey2023}; \citealt{Robertson2023}; \citealt{McLeod2024}). Most recent unobscured SFRD constraints—derived from UV luminosity function (LF) studies with JWST—show evidence of gradual evolution over the redshift range $z = 8 - 13$ (e.g., \citealt{Donnan2023a,Donnan2023b}; \citealt{Harikane2023}; \citealt{Bouwens2023a,Bouwens2023b}; \citealt{Adams2023}; \citealt{Leung2023}; \citealt{McLeod2024}). The lack of strong evolution in the bright end of the luminosity function at $z > 12.5$ (\citealt{Harikane2023}; \citealt{McLeod2024}; \citealt{Robertson2023}; \citealt{Donnan2024}) has led to suggestions that galaxies at such early epochs might experience increased star formation efficiencies (e.g., \citealt{Harikane2023}, \citealt{Yung2023}), possibly due to the lack of feedback (\citealt{Dekel2023}). However, more recent constraints at $z=12-14.5$ (\citealt{Donnan2024}) seem to be consistent with theoretical predictions (from analytical models: \citealt{Ferrara2023}; semi-analytical models: \citealt{Mauerhofer2023}, and simulations: FLARES, \citealt{Vijayan2020}; \citealt{Lovell2020}; \citealt{Wilkins2023}) without requiring major evolution in dust content and/or star formation efficiencies of massive and early galaxies.

It is important to recognize that JWST's selection of rest-frame UV-bright objects introduces a bias in our view of the SFRD at $z > 4$, where previously mentioned studies (\citealt{Gruppioni2020, Casey2021, Loiacono2021, Algera2023, Barrufet2023}) have shown that the obscured SFR can contribute significantly to the cosmic SFR density. %Complementary observations in the sub-mm are still fundamental to gather a comprehensive view of galaxy evolution from Cosmic Noon out to the Epoch of Reionization

% Spatial resolution is key to (i) resolving sources (avoiding source confusion) and (ii) correctly identifying counterparts in X-ray (especially) and optical, so that we can figure out how much is starlight and how much is accretion powered.
% Need sub-arcsecond resolution.

In the following sections of this paper, we aim to demonstrate how a survey performed by a FIR interferometer, achieving sub-arcsecond resolution, can overcome these limitations and advance our understanding of galaxy-{BH} co-evolution.

This progress will be realized by exploring the rich array of spectral features in the FIR, which hold the key to deciphering the underlying physical processes governing galaxy and AGN evolution.

Important mid/far-infrared (MIR/FIR) lines can be detected at redshifts of interest, revealing properties of the interstellar medium and the hardness of the primary radiation  (\citealt{CarilliWalter2013}, see also \citealt{Farrah2007,Farrah2013} for the main MIR/FIR spectroscopic diagnostics).

Prominent polycyclic aromatic hydrocarbon (PAH) features, which contribute a significant fraction of the total IR emission and trace the SFR \citep{Smith2007,Riechers2013,Shivaei2024}, will be visible up to high redshifts \citep{Li2020}. PAH features' ratios have been proposed as an indicator for detecting deeply dust-enshrouded AGN  \citep{Garcia-Bernete2022}. The rotational mid-IR $H_2$ emission lines allow us to estimate temperatures and masses of moderately warm gas \citep{Rigopoulou2002}.

Lines like [CII] at $157.7\,\mu$m and [OI] at $63.18$ and $145.52\,\mu$m dominate the cooling of neutral gas. The [CII] line -- which is also emitted by photodissociating regions (PDRs, see \citealt{Wolfire2022} for a recent review) -- is one of the brightest FIR lines and thus has been observed in a large number of star-forming galaxies out to $z\sim 5-8$ (see e.g. \citealt{Bethermin2020,Carniani2020,Bakx2020,Bouwens2022}). Gas in the ionized phase is traced by [OIII] at $51.81$ and $88.36\,\mu$m, and by [NII] at $121.9\,\mu$m. Thus a spectroscopic survey over the wavelength range 25$-$400\,$\mu$m, with complementary ALMA observations at longer wavelengths, can disentangle the contributions of the various ISM phases to redshifts stretching back to the epoch of reionization.

Lines with high ionization potential, like [OIV] at $25.89\,\mu$m, are excited by hard UV emission and thus trace AGN activity \citep[e.g.,][]{Sturm2002}. Because the line intensity is linearly correlated with AGN luminosity \citep{Melendez2008, Bonato2014b}, it provides an extinction-free measure of the accretion rate.

Other important AGN tracers are [NeV] lines at 14.32 and $24.31\,\mu$m  \citep[ionization potential of $\simeq 100\,$eV;][]{Tommasin2010}, and their ratio is a measure of the electron density in highly ionized regions \citep{Sturm2002, Tommasin2010}. The simultaneous detection of lines that trace the star formation and the accretion rates will allow testing of galaxy-AGN co-evolution.

An illustrative subsample of the {brightest redshifted spectral lines in the 25$-$400$\mu$m FIR spectral window}, used for the predictions provided in this paper, {is} listed in Table\,\ref{tab:lines}.

%%%%%%%
% TABLE %
%%%%%%%%
\begin{table}%[h]%
%\vskip-2.5cm
\caption{Bright spectral lines used in this paper.}
\label{tab:lines}
\centering
\footnotesize
\begin{tabular}{lc}
\hline
\hline
%\rule[-3mm]{0mm}{0mm}
Species & Wavelength [$\mu$m] \\
\hline
\multicolumn{2}{l}{\textit{Photodissociation Region}}\\
${\rm PAH}$ & 3.3, 6.2, 7.7, 8.6, 11.3, 12.7\\
${\rm H_{2}}$ & 6.91, 9.66, 12.28, 17.03 \\
${\rm [ClII]}$ & 14.38\\
${\rm [FeII]}$ & 25.98 \\
${\rm [SIII]}$ & 33.48\\
${\rm [SiII]	}$ & 34.82\\
${\rm [OI]}$ & 63.18, 145.52 \\
${\rm [CII]}$ & 157.7 \\
${\rm [CI]}$ & 370.42\\
\hline
\multicolumn{2}{l}{\textit{Stellar/HII regions}}\\
${\rm [ArII]}$ & 6.98 \\
${\rm [ArIII]}$ & 8.99, 21.82 \\
${\rm [SIV]}$ & 10.49 \\
${\rm HI}$ & 12.37 \\
${\rm [NeII]}$ & 12.81 \\
${\rm [NeIII]}$ & 15.55 \\
${\rm [FeII]}$ & 17.93 \\
${\rm [SIII]}$ & 18.71 \\
${\rm [FeIII]}$ & 22.90 \\
${\rm [OIII]}$ & 51.81, 88.36 \\
${\rm [NIII]}$ & 57.32 \\
${\rm [NII]}$ & 121.90, 205.18 \\
\hline
\multicolumn{2}{l}{\textit{AGN}}\\
${\rm [NeVI]}$ & 7.63\\
${\rm [ArV]}$ & 7.90, 13.09\\
${\rm [MgV]}$ & 13.50\\
${\rm [NeV]}$ & 14.32, 24.31\\
${\rm [OIV]}$ & 25.89\\
\hline
\multicolumn{2}{l}{\textit{Coronal regions}}\\
${\rm [SiVII]}$ & 6.50\\
${\rm [CaV]}$ & 11.48\\
\hline
\multicolumn{2}{l}{\footnotesize{In this Table, the four classes in which the spectral lines are }}\\
\multicolumn{2}{l}{\footnotesize{subdivided correspond to the regions where they are mainly }}\\
\multicolumn{2}{l}
{\footnotesize{produced. Note that some lines can be observed in multiple }}\\
\multicolumn{2}{l}
{\footnotesize{regions, albeit with varying intensities. For the subdivision }}\\
\multicolumn{2}{l}
{\footnotesize{of most of the lines we followed the \citet{Spinoglio2012} }}\\
\multicolumn{2}{l}
{\footnotesize{critical density for collisional de-excitation vs. ionization }}\\
\multicolumn{2}{l}
{\footnotesize{potential diagnostics
(their Fig.4)}}\\
\hline
\hline
\end{tabular}
\end{table}

The structure of the paper is as follows. Section\,\ref{sect:confusion} describes simulations demonstrating the importance of high resolution for beating the confusion noise, obtaining sensitive measurements of individual galaxy spectra, and exploiting emission lines to characterize physical conditions of galaxies. Section\,\ref{sect:survey} describes a case study for a reference FIR survey. It details the physically-grounded model used for our calculations and presents the related achievable results. Our results are discussed in Section\,\ref{sect:discussion}. The main conclusions from our analysis are summarized in Section\,\ref{sect:conclusions}.

Calculations are carried out adopting a standard flat $\Lambda$CDM cosmology with $\Omega_{\rm m} = 0.31$, $\Omega_{\Lambda} = 0.69$ and $h=H_0/100\, \rm km\,s^{-1}\,Mpc^{-1} = 0.677$ \citep{PlanckCollaboration2020parameters}.

\section{Probing beyond the confusion noise floor}\label{sect:confusion}

Spatial confusion is an impediment to scientific information retrieval, but how limiting is this effect? A cold ($\sim$4 K) FIR telescope equipped with state-of-the-art detectors theoretically could be limited in sensitivity only by photon noise from the natural sky background. However, a 3.5-m telescope like \textit{Herschel} barely resolves point sources separated by 7 arcseconds at $100\,\mu$m and already at flux densities well above the achievable sensitivity galaxies are much more closely spaced. For \textit{Herschel} the rms confusion noise floor was 5.8, 6.3, and 6.8 mJy/beam at 250, 350, and
$500\,\mu$m, respectively \citep{Nguyen2010}. \textit{Herschel}/SPIRE was used to measure galaxy number counts at these wavelengths down to tens of mJy \citep{oliver2010, Clements2010}. A 3.5-m or smaller telescope might survey the entire sky to the confusion limit and yet provide limited information on individual distant galaxies and galaxy evolution. In this section we use a model of the FIR sky at high resolution to assess the information lost due to spatial confusion. For the sake of illustration, we compare the spectra of individual simulated galaxies with one predicted to be seen with a cold 2-m telescope.

{We updated the previously unpublished code designed to simulate the FIR sky at high angular resolution. The code uses the best available foreground models and generates synthetic extragalactic sources in a manner consistent with \textit{Herschel} and other available measurements.} To account for the FIR foregrounds, our code uses the COBE zodiacal emission model \citep{kelsall1998}, IRAS measurements of the spatial distribution of Galactic cirrus \citep{schlegel1998}, a parametric model of the cirrus spectrum \citep{zubko2004}, and a power law with index $-2.5$ fitted to cirrus spatial structure seen with IRAS on arcminute and coarser scales to extrapolate the structure to never-before-seen sub-arcsecond scales. This foreground model estimates the specific intensity of the FIR emission as a function of sky coordinates, solar elongation angle, and wavelength.

Star-forming disk galaxies are added to the foreground emission. Empirical full-sky galaxy counts binned by flux density, wavelength, and redshift based on \textit{Herschel} observations (\citealt{Bonato2014b}) are used as a three-dimensional probability distribution function to assign random flux density and redshift values to the individual galaxies for a chosen wavelength. {Initial sky} coordinates and orientation parameters are also chosen at random, {and we adjust the initial coordinates to allow for stable clustering (\citealt{Peebles1980}). The} number of simulated galaxies is {determined by} the modeled field size {and a 1\,$\mu$Jy flux limit.} Each galaxy has a \citet{Sersic1968} brightness profile with $m=1$ \citep{Ciotti1999} and a disk half-light radius $R_{e}$ based on its luminosity and redshift. The latter relationship is derived from relations between effective radius and stellar mass \citep{Shen2003}, star formation rate and stellar mass \citep{Peng2010}, and the star formation-$L_{IR}$ relation from \citet{Clemens2013}. Size evolution at $z<1.5$ is based on the findings of \citet{vanderWel2014}, and galaxies at $z\ge1.5$ are assumed to follow the same relationship between $R_{e}$ and $L_{IR}$ as those at $z=1.5$. The model does not account for radiative transfer effects but is sufficiently realistic for the purpose at hand.

An unresolved active nucleus (AGN) lies at the center of each simulated galaxy. Its luminosity and type are assigned probabilistically based on the galaxy's spatially and spectrally integrated infrared luminosity $L_{IR}$ \citep{Vasudevan2007, Chen2013, Cai2013, Bonato2014}. The ratio of Type 1 to Type 2 AGN is based on \citet{Bianchi2012} with reference to data from \citet{Hasinger2008}. The AGN-to-star formation ratio varies from galaxy to galaxy according to the relation between star formation and accretion rate derived by \citet{Chen2013}, allowing for the observed dispersion.

\begin{figure*}
\begin{center}
\begin{tabular}{c}
\includegraphics[height=8.0cm]{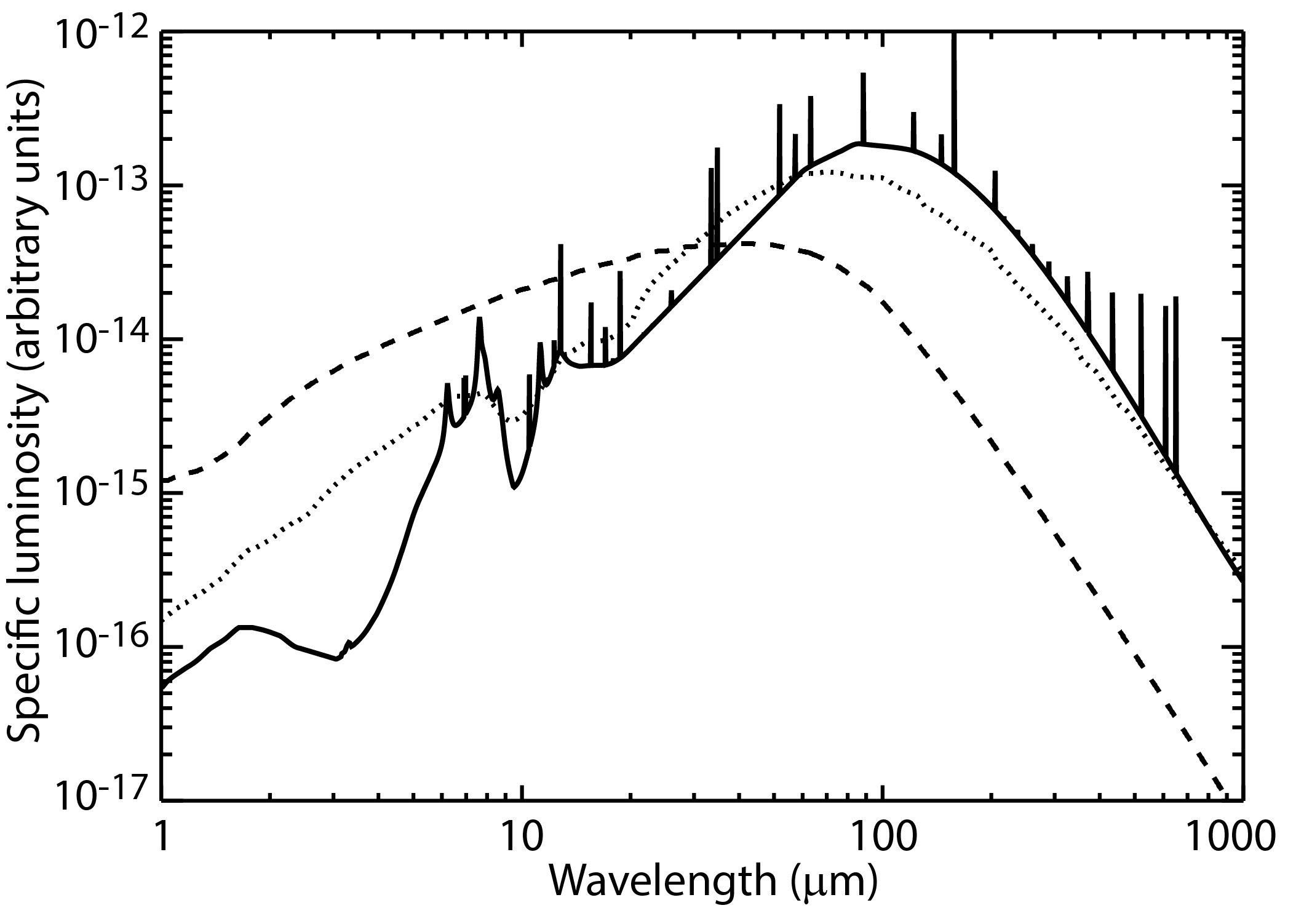}
\end{tabular}
\end{center}
\caption
{\label{fig:spectral_templates}
The same rest-frame spectra are adopted for all galaxies and AGN. The solid curve is the spectrum of a star-forming disk galaxy. The dotted curve is the adopted SED of a Type-2 AGN, and the dashed curve is the SED of a Type-1 AGN. The spectral templates are normalized such that the $8-1000\,\mu$m IR luminosity is 1 erg s$^{-1}$. Each galaxy’s spectrum is scaled to its actual luminosity. {These SEDs are taken from \citet{Cai2013}.}
}
\end{figure*}

Rest-frame spectra are adopted for the disk emission and for each of the two AGN types {\citep{Cai2013, Bonato2019}}. In a step toward simplicity, we use the same rest-frame template spectra for all galaxies and AGN and we model only the spectral energy distributions of AGN, ignoring their line emission
(Fig.\,\ref{fig:spectral_templates}). Thus, evolutionary effects and, e.g., the effects of star formation rate on a galaxy's emission will not be seen in the simulated galaxy spectra.

\begin{figure*}
\begin{center}
\begin{tabular}{c}
\includegraphics[height=8.0cm]{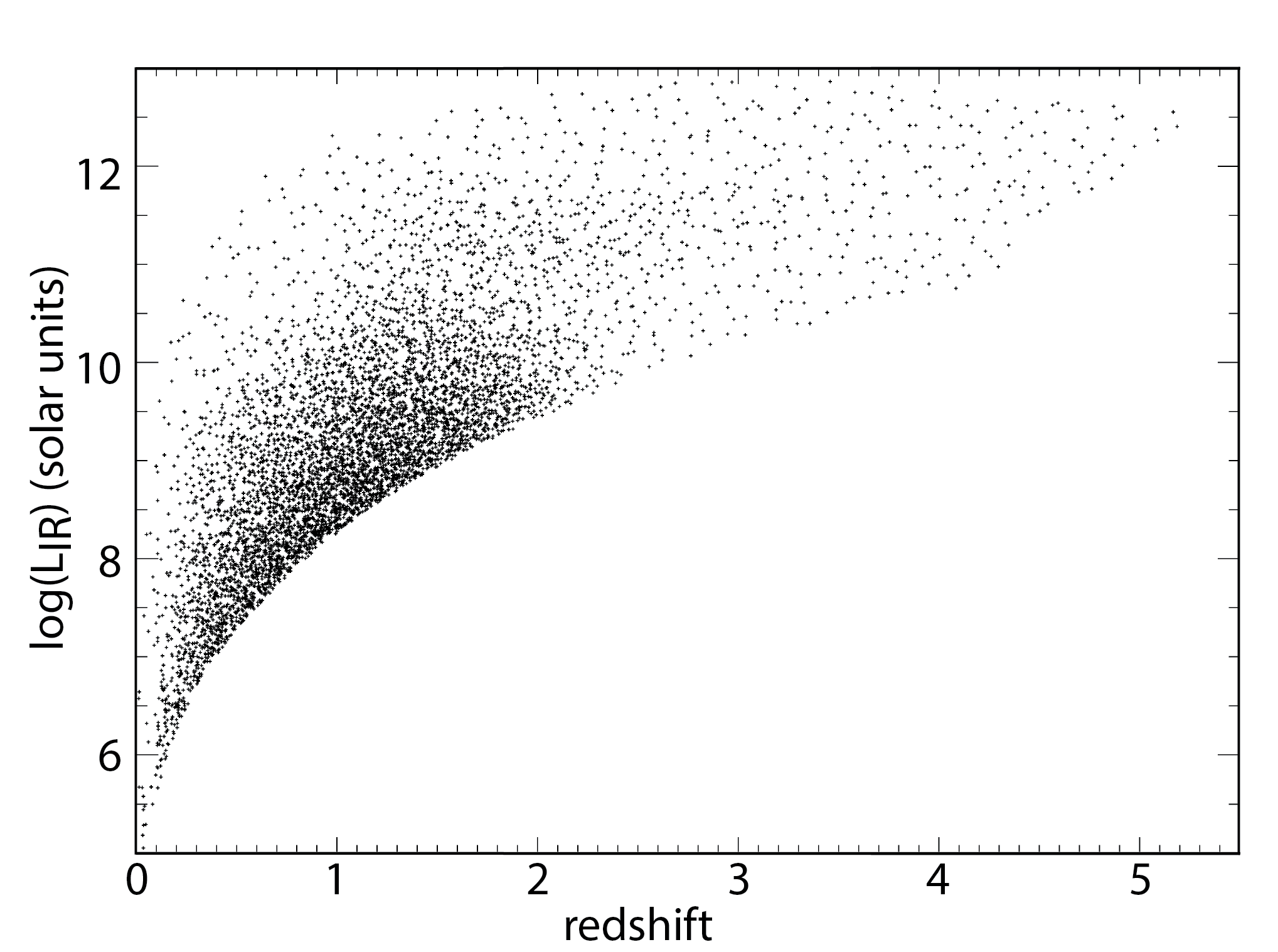}
\end{tabular}
\end{center}
\caption
{\label{fig:LvZ}
{Total} infrared luminosities (sum of star-forming disk and AGN {components}) and redshifts of 6275 galaxies in the simulated COSMOS field area. {The lower envelope corresponds} to the fact that source counts were modeled to a $1\,\mu$Jy lower bound in flux density.
}
\end{figure*}

We modeled a $4.74 \times 4.74$ arcmin field at the Galactic coordinates (l,b) = $236.^{\circ}822$, $42. ^{\circ}1216$, chosen for its location in the COSMOS deep field \citep{scoville2007}. The simulated field was assumed to be observed at a solar elongation angle of $160^{\circ}$, which affects the zodiacal emission brightness. Figure\,\ref{fig:LvZ} shows the redshifts and derived luminosities of the 6275 galaxies in the simulated field. {The mean separation between nearest neighbors is approximately 1.8 arcseconds.}

We calculated model maps of specific intensity at wavelengths ranging from $25\,\mu$m to $400\,\mu$m in $0.05\,\mu$m increments at wavelengths $\le100\,\mu$m and $0.5\,\mu$m increments at longer wavelengths. Collectively, the synthesized images comprise a hyperspectral data cube with 2101 wavelength slices. Each image has 8192$^2$ pixels at 0.0347’’/pixel, enabling convolution to the resolution of an interferometer or a single-aperture telescope. Spectral lines are smoothed to $\Delta \lambda = \lambda\,(\mu\hbox{m})/300$ wavelength channels to derive equivalent flux densities.

\begin{figure*}
\begin{center}
\begin{tabular}{c}
\includegraphics[height=20.0cm]{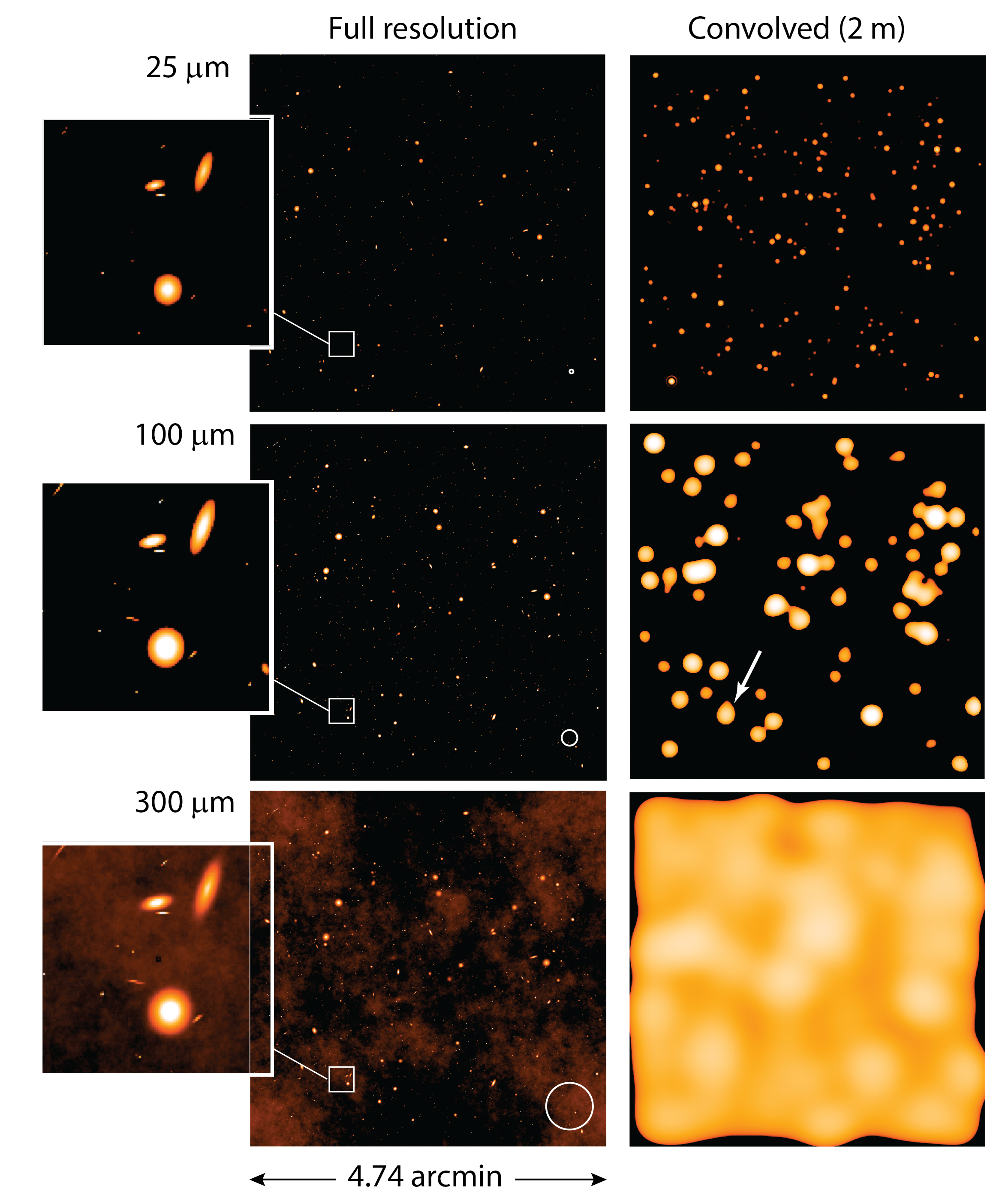}
\end{tabular}
\end{center}
\caption
{\label{fig:SimSkyImages}
25, 100, and 300-$\mu$m wavelength slices from the simulated hyperspectral data cube (left with insets that reveal the richness of the model dataset) and convolved to the resolution of a 2-m telescope (right). The beam size of a 2-m telescope at each wavelength is shown with a circle, indicating that light from most or all the galaxies in the inset area would blend at 100$\,\mu$m and longer wavelengths. A white arrow in the 100-$\mu$m convolved image points to the location of the aperture analyzed in the spectral domain.
}
\end{figure*}

Figure\,\ref{fig:SimSkyImages} shows 25, 100, and 300-$\mu$m slices from the simulated data cube and corresponding images convolved to the resolution of a notional 2-m diameter diffraction-limited telescope with a 20-cm central obscuration. Each image is shown on a logarithmic intensity scale with minima close to the foreground intensities at each wavelength and maxima of 50 MJy/sr in all the full-resolution maps and 50, 10, and 5 MJy/sr in the convolved maps at 25, 100, and 300$\,\mu$m, respectively. Cirrus emission is noticeable in the 300-$\mu$m full-resolution image. We ignore photon shot noise. {Due to beam dilution, the faintest galaxies are not visible above the foreground emission in the 2$-$m telescope images,} even at $25\,\mu$m {where confusion is negligible.}

\begin{figure*}
\begin{center}
\begin{tabular}{c}
\includegraphics[height=12.0cm]{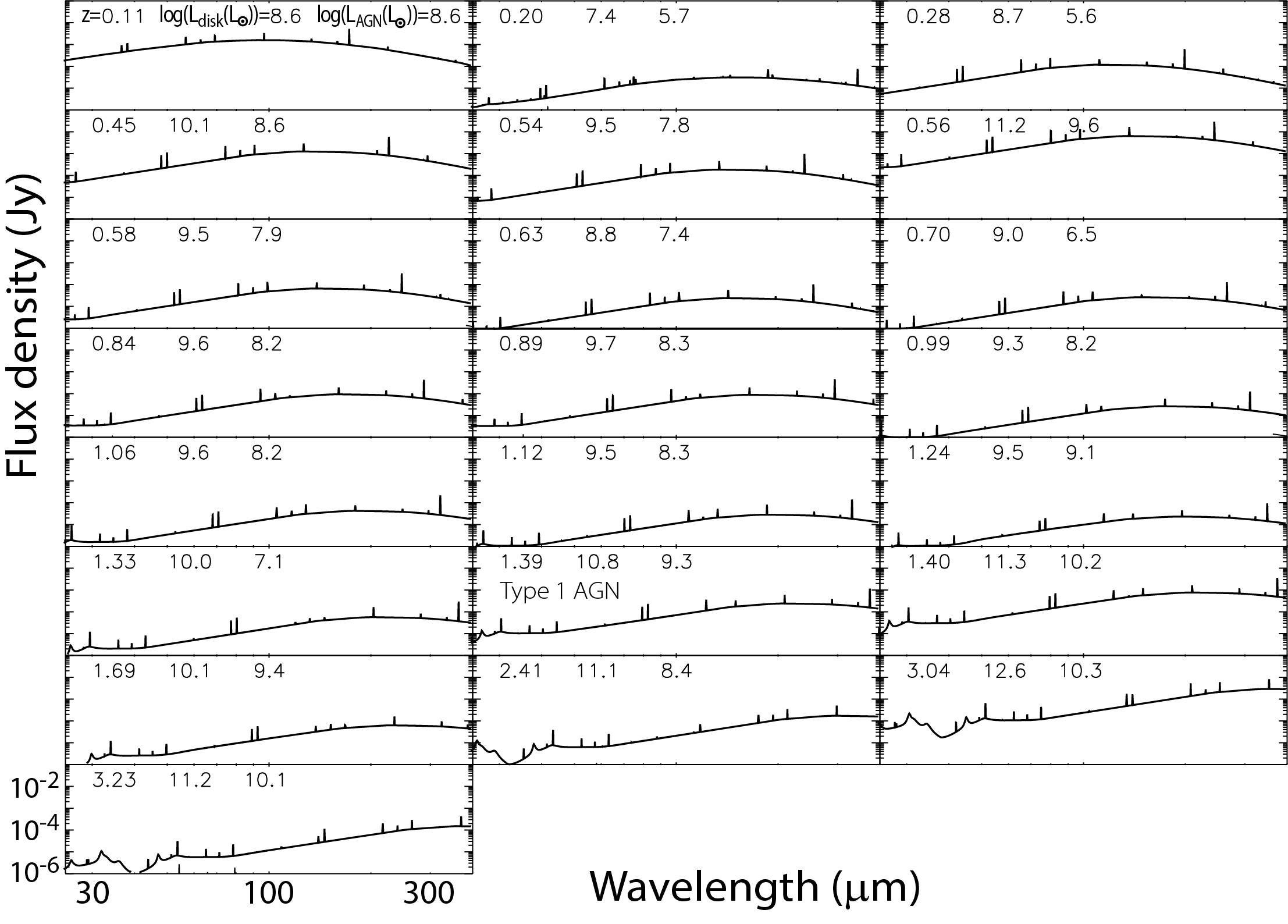}
\end{tabular}
\end{center}
\caption
{\label{fig: spectra_of_galaxies_in_Aperture_2}
Spatially integrated spectra of the 22 brightest of 48 simulated galaxies in the analyzed aperture area {shown in order of increasing redshift in raster fashion from top-left to bottom-right}. The remaining galaxies are at best barely brighter than 10$\,\mu$Jy. The same foreground spectrum was subtracted from each galaxy’s spectrum. The foreground spectrum is extracted from the data cube at the location of the darkest pixel at $25\,\mu$m. Three numbers in each panel indicate the galaxy’s redshift, the base-ten logarithm of its {disk-integrated} infrared ($8-1000\,\mu$m) luminosity in solar units, and the galaxy’s AGN luminosity in the same units, respectively. Only one bright galaxy, as indicated, has a Type 1 AGN, while all the others have Type 2 AGN. The galaxy at redshift $z=3.04$ is the only one in the aperture area with a luminosity greater than $10^{12}\,L_{\Sun}$, and would be considered ``ultraluminous.''
}
\end{figure*}

We probed the spectral domain to understand more fully the effects of confusion, choosing for analysis an aperture centered on the amorphous bright spot at the tip of the white arrow in Fig.\,\ref{fig:SimSkyImages}. The aperture area corresponds to the 100-$\mu$m Airy disk of a 2-m telescope - $2.44 \lambda/d$ - and the same aperture was used at all wavelengths. Forty-eight simulated galaxies lie within the aperture. The spectra of 22 of these galaxies - those with significant emission greater than $10\,\mu$Jy - are shown in Fig.\,\ref{fig: spectra_of_galaxies_in_Aperture_2}. Two of the 48 galaxies host a Type-1 AGN; the rest are of Type 2. None of the AGN outshine their host galaxies, whereas that happens in a handful of cases in the overall sample of 6275 simulated galaxies. If they could be measured, each of these spectra would have a story to tell about the galaxies’ physical conditions, metallicity, and importance of the AGN, as well as the redshift. Collectively such spectra would inform our understanding of galaxy evolution and the co-evolution of the galaxies and their central supermassive black holes (SMBHs). An observatory offering sub-arcsecond angular resolution could measure the FIR spectra of thousands of galaxies in a field the size of that shown in Fig.\,\ref{fig:SimSkyImages}. Several fields could be observed to allow for cosmic variance.

\begin{figure*}
\begin{center}
\begin{tabular}{c}
\includegraphics[height=10.0cm]{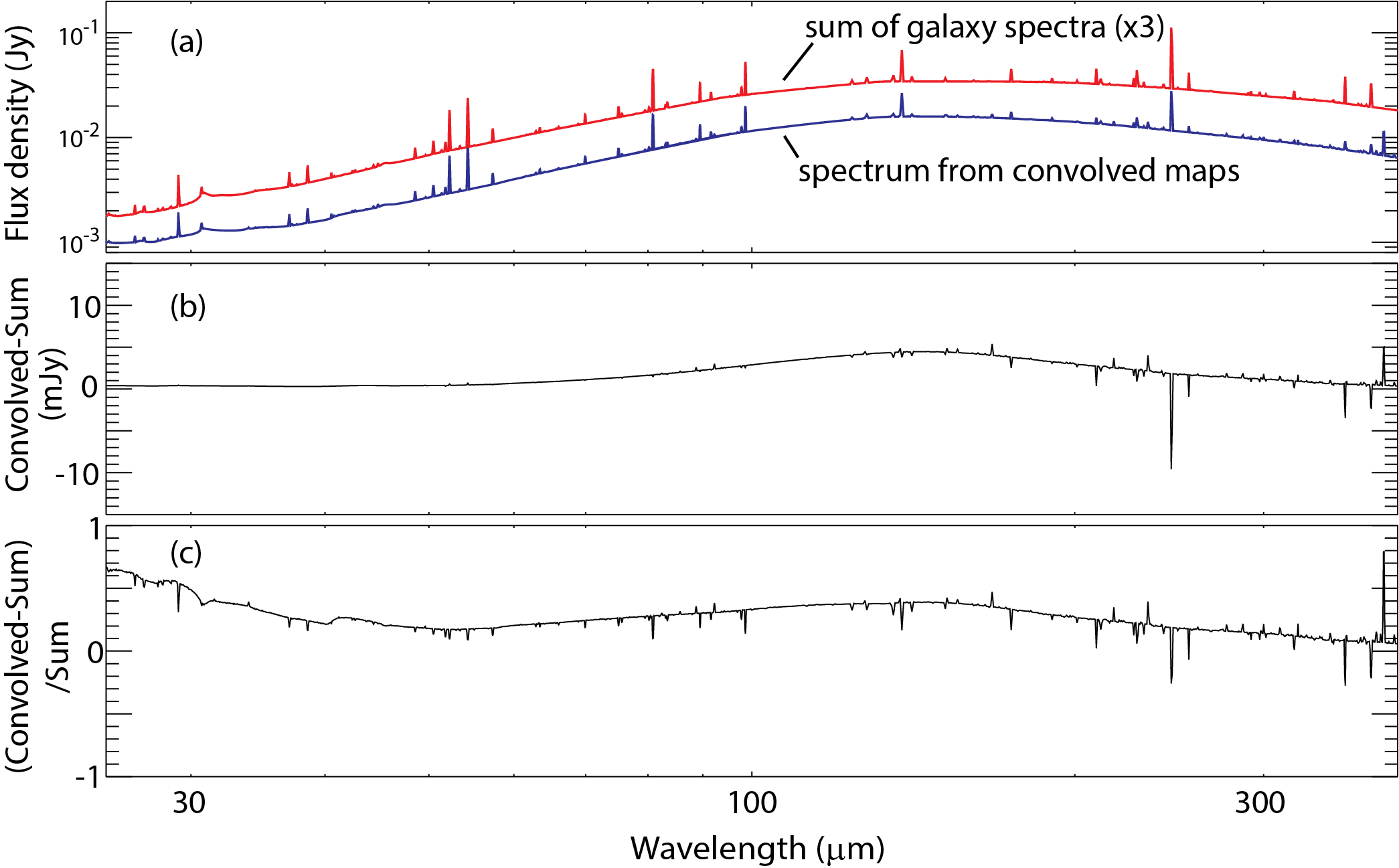}
\end{tabular}
\end{center}
\caption
{\label{fig: convolved_spectrum}
(a) (blue) Spectrum extracted from the convolved maps integrated over the same aperture area as that containing 48 galaxies, of which spectra of the brightest sources are shown in Figure\,\ref{fig: spectra_of_galaxies_in_Aperture_2}, with the ``known'' model foreground emission subtracted. (red) The summed spectrum of all 48 galaxies in the aperture {multiplied by a factor of three} to offset the resulting spectrum from the lower curve. Six galaxies - those at redshifts 0.11, 0.45, 0.56, 1.39, 1.40, and 3.04 $-$ contribute significantly to the emission seen in the summed spectrum, and a single galaxy at z$=$0.56 dominates. The PAH feature seen at $\sim$30\,$\mu$m is primarily attributable to the ultraluminous galaxy at z$=$3.04. (b) Difference between the convolved spectrum (lower/blue curve in (a)) and the sum of the galaxy spectra in the analyzed aperture (upper/red curve in (a)), and (c) the corresponding fractional difference between the two spectra. Spectral line blending is most severe at the long-wavelength end of the spectrum due to the larger beam size at longer wavelengths. Line intensities can be either underestimated or overestimated in a spectrum obtained with a 2$-$m telescope. At wavelengths $\gtrsim$70\,$\mu$m, the shape of the continuum is distorted, suggesting that efforts to ascribe continuum emission to individual galaxies are subject to bias from sources outside the aperture area. While not apparent in (b), the fractional difference (c) shows that broadband PAH features from high-redshift galaxies merge at wavelengths $\lesssim$50\,$\mu$m.}
\end{figure*}

What would an on-axis 2-m diameter telescope see? To answer that question we integrate spatially over the pixels in the same aperture area as that described above but derive the spectrum from convolved images like those shown on the right-hand side of Fig.\,\ref{fig:SimSkyImages}. The spectrum derived from the convolved images is shown in the lower curve in Fig.\,\ref{fig: convolved_spectrum}-(a) panel. Because even the darkest pixel in the 25-$\mu$m convolved map contains significant extragalactic source emission, a model is needed to subtract the foreground spectrum. Here we are fortunate to know the computed foreground, but an observer who encounters confusion noise will only have an estimate, leaving room for systematic error in the foreground subtraction. Uncertainty in the foreground could affect the continuum shape in the aperture-integrated spectrum.

For comparison with the spectrum derived from the convolved maps, the upper curve in Fig.\,\ref{fig: convolved_spectrum}-(a) panel shows the sum of the spectra of all the individual galaxies in the aperture area {(multiplied by a factor of three to produce an offset for clarity)}. Panels (b) and (c) of Figure\,\ref{fig: convolved_spectrum} show the difference between the convolved-map spectrum and the summed galaxy spectra and indicates that both continuum and line emission seen in the convolved maps are not a simple mixture of the spectra of the individual galaxies. Line emission can be underestimated or overestimated. The shape of the continuum seen with a 2-m telescope is not simply the sum of the continua of the blended galaxies because the telescope's beam size is wavelength-dependent. {If we had} analyzed a larger aperture area, we would have found smaller differences, but then {the} emission from a greater number of galaxies would be blended.

Sophisticated retrieval techniques can be used to probe below the classical confusion noise floor by a factor of $\sim$3 (\citealt{Hurley2017}, \citealt{Donnellan2024}), and it is possible to derive redshifts of the bright galaxies that dominate the spectrum seen with a 2-m telescope. However, blending in the spectral domain and the lack of information about each galaxy’s continuum emission compromise the derivation of line intensities. As described in Sect.\,\ref{sect:introduction}, a great deal of information, such as information about metallicity, physical conditions in the interstellar medium, and the presence or absence of a significant AGN, is contained in FIR spectral lines and their intensity ratios. Without accurate line intensities much of this information would be lost or plagued with systematic uncertainty.

{The} convolved-map spectrum differs from the sum of the spectra of the 48 galaxies in the aperture area chosen for analysis because the beam of a 2-m telescope encompasses emission outside the 100-$\mu$m Airy disk at longer wavelengths. We find significant spectrally complex differences between the two spectra shown in Fig.\,\ref{fig: convolved_spectrum}-(a) panel, illustrating the extent to which information in the spectral domain is lost or skewed as a consequence of spatial blending.

To ensure that these results are not anomalous, we repeated the analysis for another aperture centered on a different, somewhat less isolated bright spot in the convolved 100-$\mu$m image. The results were  qualitatively similar to those shown in Figure \ref{fig: convolved_spectrum}.

Six galaxies dominate the spectrum shown in the lower curve in Fig.\,\ref{fig: convolved_spectrum}-(a) panel. These galaxies lie in the redshift range 0.11 to 3.04 and half of them would be classified as Luminous Infrared Galaxies (LIRGs) with $L_{IR} > 10^{11}\, L_\odot$. Similar results were found in the second analyzed aperture. This population is not representative of the general galaxy population.

{With sufficient spectral resolution one could discern the presence of multiple bright galaxies in a 2$-$m telescope’s beam, count those galaxies, and measure their redshifts, but many faint galaxies would be overlooked. A model could be used to estimate how much continuum emission to assign to each discernible galaxy as a function of wavelength, and such a model could allow for possible evolution in galaxy luminosity, dependence on star formation rate and dust filling fraction, metallicity, etc., all unknown and indeed some of the parameters one would like to measure. However, assigning a portion of the continuum to each of the contributing galaxies is fraught with uncertainties and potential model degeneracies. At best, one might find a self-consistent solution and have little insight into systematic errors.}

{To probe the co-evolution of galaxies and their central SMBHs and learn how galaxies formed and built up heavy elements and dust over cosmic time, major goals of the Astro2020 Decadal Survey \citep{decadalsurvey2021}, we need sensitive far-IR spectroscopic observations of galaxies out to high redshifts. We must characterize physical conditions, such as the hardness of the interstellar radiation field, measure metallicity, and test hypotheses for how these galaxy characteristics change as a function of redshift. FIR spectral line strength ratios, PAH features, and the continuum shape all contain the desired information. Thus, direct measurements of individual galaxy spectra are far superior probes of evolution. Such measurements are possible with an observatory that offers sub$-$arcsecond resolution, spectral resolving power in the hundreds to thousands, and sensitivity of the order of micro$-$Janskys.}

% In this paper we discuss possible SPICE surveys and present quantitative predictions for counts in the continuum and in the main emission lines. We adopt as our reference a shallow survey of 1000\,hr (including overheads; $\sim$897.4\,h on source), covering $\sim0.56\,\hbox{deg}^2$ and reaching a $5\,\sigma$ detection limit of $\sim$0.27\,mJy in the continuum. Section\,\ref{sect:model} contains a description of the model while the results are presented in Sect.\,\ref{sect:results}, where we also explain the reason for the choice of the reference survey. The main conclusions are summarized and discussed in Sect.~\ref{sect:conclusions}. Calculations are carried out adopting a standard flat $\Lambda$CDM cosmology with $\Omega_{\rm m} = 0.31$, $\Omega_{\Lambda} = 0.69$ and $h=H_0/100\, \rm km\,s^{-1}\,Mpc^{-1} = 0.677$ \citep{PlanckCollaboration2020parameters}.

\section{Probing galaxy-AGN co-evolution with a high-resolution FIR survey}\label{sect:survey}

\begin{figure*}
\begin{center}
\begin{tabular}{c}
\includegraphics[height=5.5cm, width=3.2in]{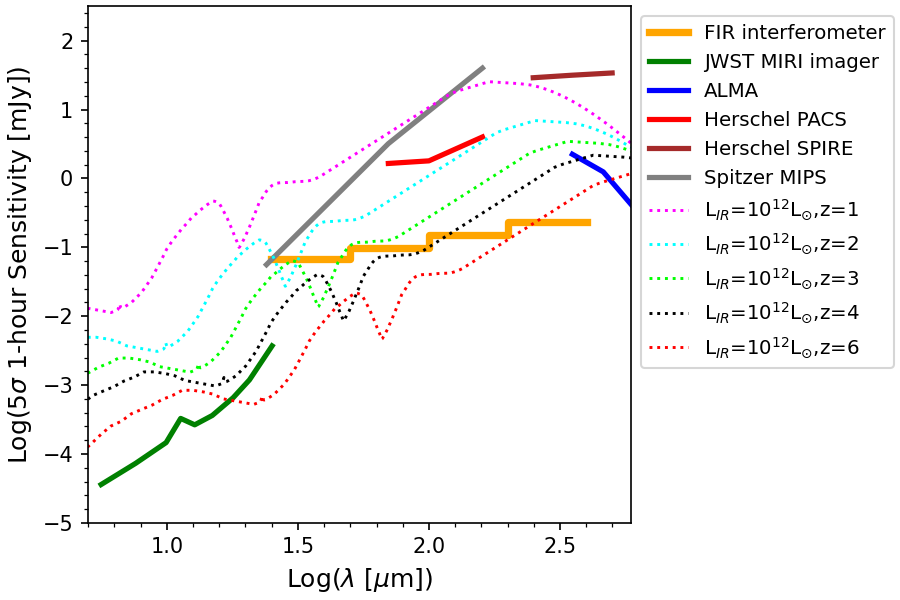}  % fig2 includes two images
\includegraphics[height=5.5cm, width=3.2in]{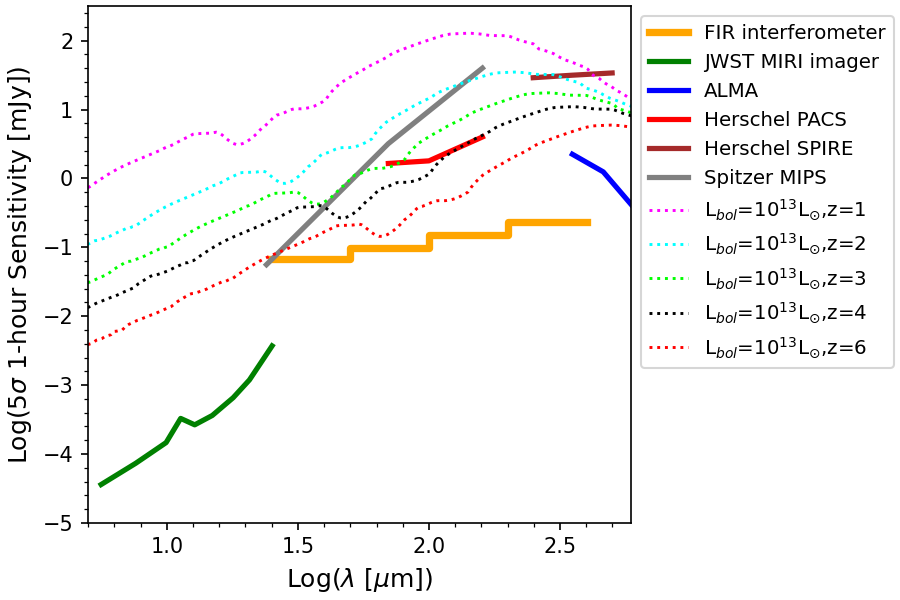}  % fig2 includes two images
%\\
%(a) \hspace{5.1cm} (b)
\end{tabular}
\end{center}
\caption
{ \label{fig:continuum}
Template SEDs of dusty proto-spheroids with $L_{\rm IR}=10^{12}\,L_\odot$ (left panel) and obscured AGN with bolometric luminosity $L_{\rm bol}=10^{13}\,L_\odot$ (right panel), at $z=1$, 2, 3, 4, 6, % (from left to right),
from \citet{Cai2013}, compared with the $5\,\sigma$ detection limits for 1\,h exposure of the FIR interferometer, of JWST MIRI, of \textit{Spitzer} MIPS, of \textit{Herschel} PACS and SPIRE, and of ALMA. The luminosity of $10^{12}\,L_\odot$ corresponds to a SFR of $\simeq 100\,M_\odot\,\hbox{yr}^{-1}$, typical of star-forming galaxies at the cosmic noon ($z\simeq 2$--3).}
\end{figure*}

\begin{figure*}
\begin{center}
\begin{tabular}{c}
\includegraphics[height=5.5cm, width=3.2in]{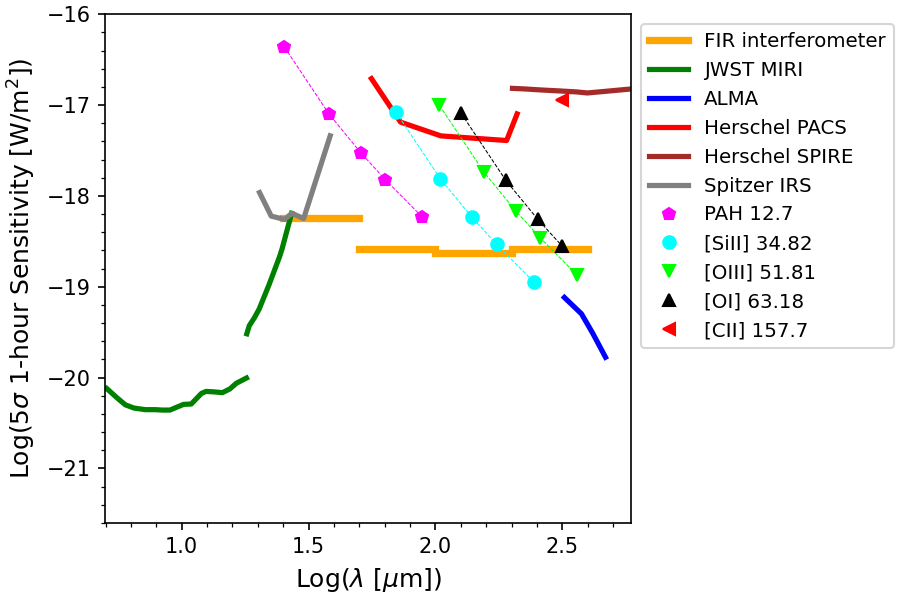} % \\  % fig2 includes two images
\includegraphics[height=5.5cm, width=3.2in]{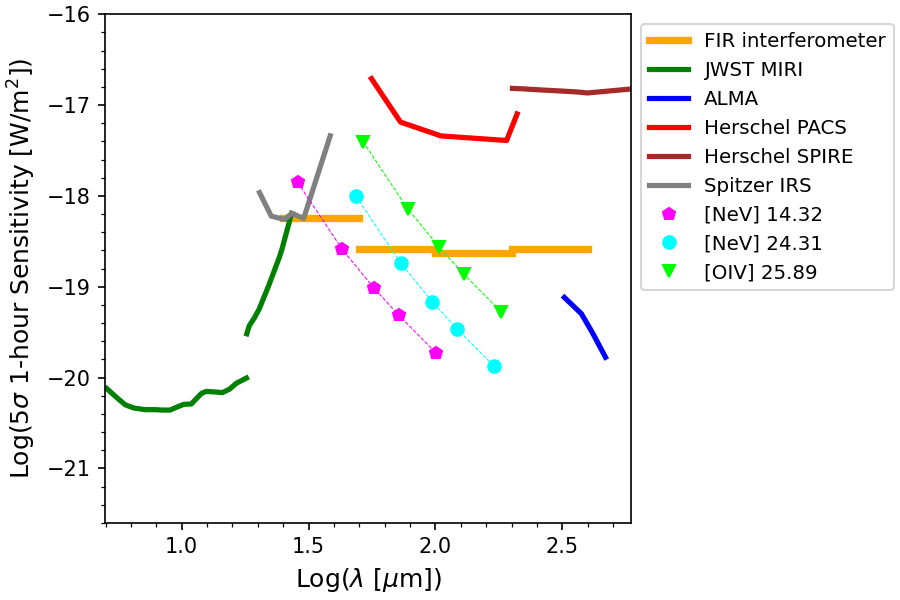}  % fig2 includes two images
%\\
%(a) \hspace{5.1cm} (b)
\end{tabular}
\end{center}
\caption
{ \label{fig:lines_FIR}
Predicted intensities of key spectral lines, compared to $5\,\sigma$ detection limits for 1\,h exposures with the FIR interferometer, {\it JWST} MIRI, \textit{Spitzer} IRS, \textit{Herschel} PACS and SPIRE, and ALMA.
\textit{Left panel:} Lines powered by star formation from galaxies with $L_{\rm IR}=10^{13}\,L_\odot$ at $z=1$, 2, 3, 4, and 6 (filled symbols connected by solid lines, from left to right).
\textit{Right panel:} Lines powered by AGNs with bolometric luminosity $L_{\rm bol}=10^{14}\,L_\odot$ at the same redshifts.}
\end{figure*}

As mentioned in Sect.\,\ref{sect:introduction}, sub-arcsecond resolution is attainable at FIR wavelengths with interferometry.
Given the significant advancements in FIR interferometry technology, a large-scale space mission in the next decade is highly feasible. Such a mission could provide invaluable insights into the co-evolution of galaxies and black holes. To demonstrate the potential of such a mission and to benchmark a notional high-resolution FIR sky survey (hereafter the ``reference survey''), we consider the measurement capabilities of the Space Infrared Interferometric Telescope (SPIRIT; \citealt{Leisawitz2007, Leisawitz2008, Leisawitz2009}), which can achieve the James Webb Space Telescope (JWST)-like resolution essential for studying the co-evolution of galaxies and accreting black holes. Referring to Table\,3 of \citet{Leisawitz2009}, SPIRIT has wavelength range 25--$400\,\mu$m; instantaneous field-of-view (FoV) 1\,arcmin; angular resolution 0.3 $(\lambda/100\,\mu\hbox{m})$ arcsec. The point source continuum sensitivities ($5\,\sigma$, 24 hours per FoV) are 14, 20, 31, and $48\,\mu$Jy, at 35, 70, 140, and $280\,\mu$m, respectively. The instrument design provides, in addition to spatial interferometry, Fourier transform spectroscopy with spectral resolution $R=3175$, 5058, 4265, and 3000, and spectral line detection limits of 2.9, 1.7, 1.4 and $1.3\times 10^{-19}\,\hbox{W}\,\hbox{m}^{-2}$ respectively, at the 4 wavelengths and under the conditions mentioned above. The spectral resolution comes from scanning an optical delay line and it is given by the number of detectable fringes above some noise threshold. Because a moderate spectral resolution of $\sim500$ is enough for the purposes discussed in this paper, we would Fourier transform the full interferogram to obtain a high-$R$ spectrum, and rebin the spectral channels, gaining a factor $\sim (R/500)^{1/2}$ in sensitivity.

Fig.\,\ref{fig:continuum} shows that such a FIR interferometer would fill the gap in our view of the cosmic star formation history mentioned in Sect.\,\ref{sect:introduction} and shown in Fig.\,\ref{fig:SFRH}. It would allow us to investigate the evolution of dust temperature (or, more directly, the SED peak temperature which is largely unconstrained at $z>2$ \citep{Sommovigo2022}) with cosmic time, which needs to be matched by evolutionary models.

On the spectroscopic side, the wavelength gap between JWST and ALMA translates into gaps in the redshift ranges over which MIR/FIR lines can be detected, as illustrated by Fig.\,\ref{fig:lines_FIR}. This limits the possibility of exploiting the emission lines to study the properties of the interstellar medium (density, temperature, ionization state), to investigate gas heating and cooling processes and the hardness of the radiation field linked to the presence of AGN (see \citealt{CarilliWalter2013} for a review). Such a FIR interferometer would be a valuable complement of ALMA and JWST.

%By resolving the rich variety of FIR spectral lines, it would be possible to probe the kinematics of the interstellar medium, providing insights into the processes driving galaxy evolution. High-resolution data would allow us to measure the velocity fields of the interstellar medium, revealing the presence of inflows, outflows, and rotation, and providing constraints on the physical processes regulating star formation and AGN feedback. By studying the kinematics of galaxies at different redshifts, we can trace the evolution of galaxy disks and bulges, and understand the role of mergers and feedback in shaping galaxy properties.

This Section describes the science achievable for a reference extragalactic survey covering a half square degree in 1000 hours (including overheads, $\sim$900\,h on sources), with predictions for the numbers of galaxies and AGN detected in the continuum and in the main emission lines. The reference survey will achieve $5\,\sigma$ point source detection limits of $(8.0, 3.7, 3.3, 3.7)\times 10^{-19}\,\hbox{W}\,\hbox{m}^{-2}$ in spectroscopy and (0.10, 0.14, 0.21, 0.33)\,mJy in the continuum, for the 25--50, 50-100, 100--200 and 200--400$\,\mu$m wavelength intervals, respectively.
This survey will detect emission from dust heated by young stars in galaxies at redshifts of up to $z\sim 7$, or even higher if the typical dust temperature continues to increase with redshift \citep{Bouwens2020,Sommovigo22, Algera23b} --- which the survey will determine. The attained angular resolution will allow us to reveal the evolving structure of dusty star-forming galaxies, which reflect the physical processes that drove the galaxy buildup, such as mergers, interactions, accretion flows, shrinking of the star-formation region by dissipation, or morphological transformations.

The reference survey angular resolution at $30\,\mu$m (0.09 arcsec) corresponds, for the adopted cosmology, to linear sizes $\le 0.78\,$kpc at $z> 1$. {The available information on the size of dusty star forming regions (or FIR sizes) in moderate- to high-redshift massive galaxies is still limited}. \citet{Rujopakarn2011} found a median diameter of 4.4\,kpc for a sample of luminous/ultra-luminous infrared galaxies and sub-millimeter galaxies (SMGs) at $0.4<z<2.7$, while \citet{Gullberg2019} derived a typical effective radius at $870\,\mu$m of $\simeq 0''.15 \pm 0''.05$ for a sample of 153 bright SMGs with a quartile range in redshift of 2.5--3.5.  \citet{Tadaki2020} reported a broad range of FIR effective radii ($0.4\,\hbox{kpc}< R_e < 6\,\hbox{kpc}$) for massive star-forming galaxies at $1.9 < z <2.6$.  {Similar results can also be found in \citet{Simpson2015, Barro2016, Simpson2017, Hodge2019, Rujopakarn2019, Lang2019, Pantoni2021} and in the review by \citet{Hodge2020}. At higher redshift, \citet{McKinney2023} provide an upper limit of 0.44'', in ALMA Band 6, for the effective radius of a FIR-luminous dusty galaxy at z$\sim$5.}%\citet{Smolcic2015}  inferred a median radio-emitting size for their six $z> 4$ SMGs of ($0''.63 \pm 0''.12) \times (0''.35 \pm 0''.05$).

The stretching of images by gravitational lensing will make it possible to extend the morphological analysis to longer wavelengths for a sizable subset of detections. The proposed survey will shed light on this debated issue, namely the sizes of star-forming regions in distant galaxies, and on the origin of the galaxies' morphological diversity.

\subsection{Model}\label{sect:model}

For the predictions provided in this Section, we use a more sophisticated modeling approach than that used in Sect.\,\ref{sect:confusion}, where only three reference SEDs for the evolution of galaxies/AGN populations were considered. We adopt an updated version of the physically grounded analytical model by \citet{Cai2013}, describing the evolution of eight different populations of galaxies/AGN. The model predicts the co-evolution of spheroidal galaxies and of the associated AGN as a function of galactic age, halo mass, and redshift. The history of the galaxy and AGN bolometric luminosities is computed by solving a set of equations describing the evolution of three gas phases. The hot gas, initially at the virial temperature, cools and falls towards the galaxy center; the cold gas condenses into stars; the third phase consists of gas gathered around the super-massive black hole and accreting onto it at a rate regulated by viscous dissipation of the angular momentum. The equations include recipes for supernova and AGN feedback. The model yields, at each galactic age, the bolometric luminosity of the AGN and of the host galaxy, given the halo mass and the galaxy formation redshift. The luminosity at any frequency is derived using appropriate SEDs for the galaxy and the AGN. \citet{Cai2013} also computed the gravitational amplification distribution, i.e. the probability for a galaxy at redshift $z$ to have its flux density amplified by a factor $\mu$, as well as the clustering properties of galaxies exploiting the Halo Occupation Distribution formalism by \citet{Zheng2005}.

Stellar populations of present-day spheroidal galaxies and galactic bulges are relatively old, implying that they mostly formed at $z\gtrsim 1.5$. According to the model, star-forming spheroids (proto-spheroids) were the dominant contributors to cosmic star formation at high redshifts. Below $z\sim 1.5$, star-formation happens primarily in starburst and late-type ``normal'' galaxies. The model attributes different SEDs and different evolutionary properties to the two galaxy classes. Starburst galaxies contain warmer dust and evolve faster than ``normal'' galaxies. In both cases, phenomenological evolutionary laws are adopted. At variance with the model for spheroids, which co-evolves the stellar and AGN components,  AGNs associated with starburst and late-type galaxies were treated by \citet{Cai2013} as an independent population.

A tabulation of the adopted SEDs, of counts and of luminosity functions at many frequencies are available in the website \url{http://staff.ustc.edu.cn/~zcai/galaxy_agn/index.html} where extensive comparisons of model predictions with multi-frequency data at several redshifts are also presented. Although relatively old, the model is still highly competitive (see, e.g., \citealt{Chen2022} and  \citealt{Ward2022}).

In previous work, we upgraded the original phenomenological model in several ways \citep{Bonato2014b}.
We linked the black-hole accretion rate, hence the AGN bolometric luminosity, to the SFR, hence to the infrared  luminosity, so that the global emission of the galaxy, including the AGN contribution, is treated self-consistently.
We extended the study to IR lines excited by AGNs, working out relations between line and AGN bolometric luminosity.
In another paper, we derived relationships between the main IR lines and the star formation rate (SFR), complementing the available data with extensive simulations which took into account the effect of dust extinction \citep{Bonato2014a}.
Finally, we updated some of these relations taking into account more recent data and determined relations for additional lines \citep{Bonato2019}.

%%%%%%%%%%%%%%%%

\subsection{Results}\label{sect:results}

\begin{figure*}
\begin{center}
\begin{tabular}{c}
\includegraphics[trim=3.7cm 0.55cm 1.6cm 0.25cm,clip=true,width=0.32\textwidth, angle=0]{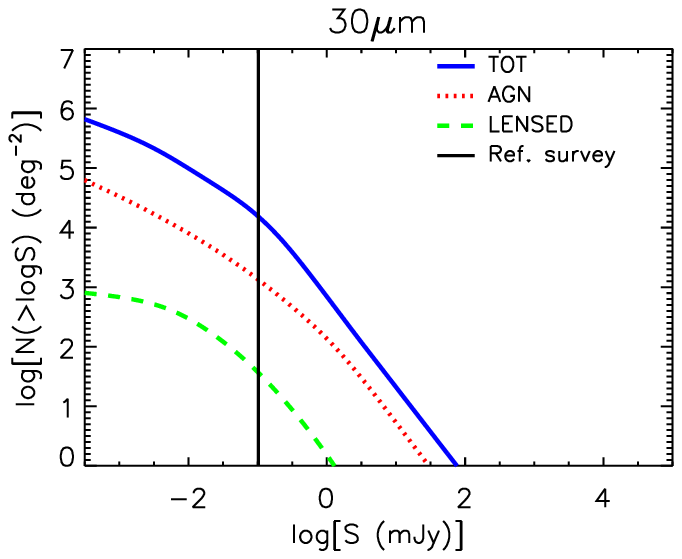}
\includegraphics[trim=3.7cm 0.55cm 1.6cm 0.25cm,clip=true,width=0.32\textwidth, angle=0]{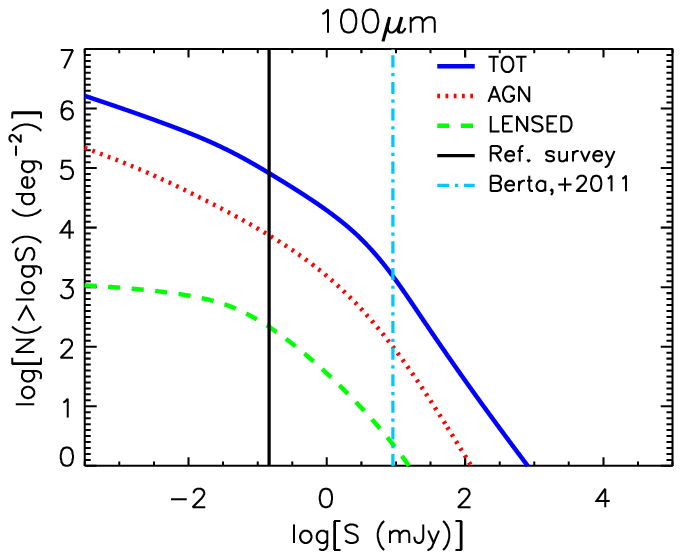}
\includegraphics[trim=3.7cm 0.55cm 1.6cm 0.25cm,clip=true,width=0.32\textwidth, angle=0]{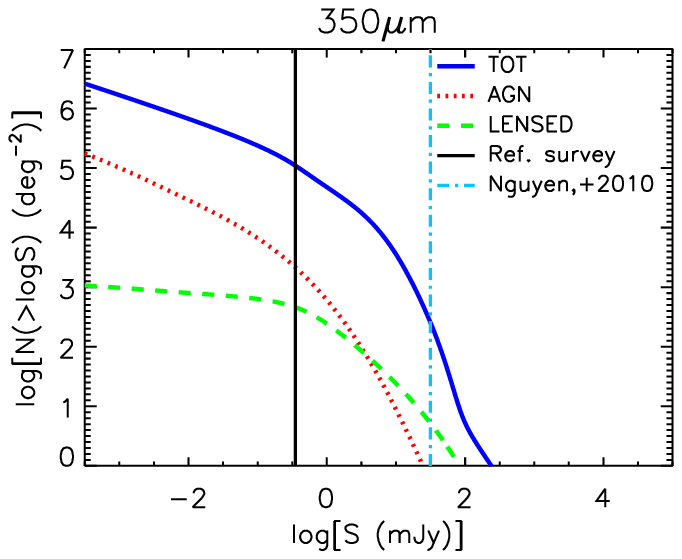}
\end{tabular}
\end{center}
\caption
{ \label{fig:counts_continuum} Integral counts per square degree at 30, 100 and $350\,\mu$m (from left to right) predicted by our model. The vertical black lines correspond to the $5\,\sigma$ reference survey detection limits of 0.14, 0.21, 0.33\,mJy, respectively.  The vertical dot-dashed cyan lines show, at $100\,\mu$m, the 80\% completeness limit (9\,mJy) of the \textit{Herschel}/PACS Evolutionary Probe (PEP) survey of the COSMOS field \citep{Berta2011}, and, at $350\,\mu$m, the \textit{Herschel}/SPIRE $5\,\sigma$ confusion noise \citep[31.5\,mJy;][]{Nguyen2010}.
 }
\end{figure*}

\begin{figure*}
\begin{center}
\begin{tabular}{c}
\includegraphics[trim=3.7cm 0.65cm 1.6cm 0.25cm,clip=true,width=0.31\textwidth, angle=0]{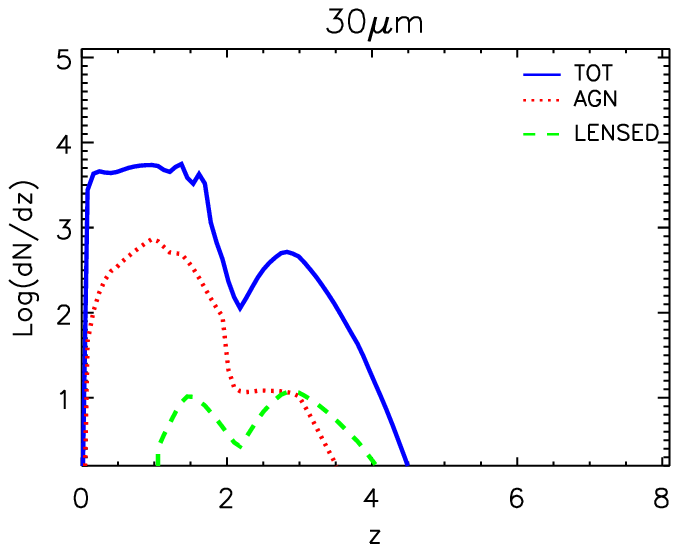}
\includegraphics[trim=3.7cm 0.65cm 1.6cm 0.25cm,clip=true,width=0.31\textwidth, angle=0]{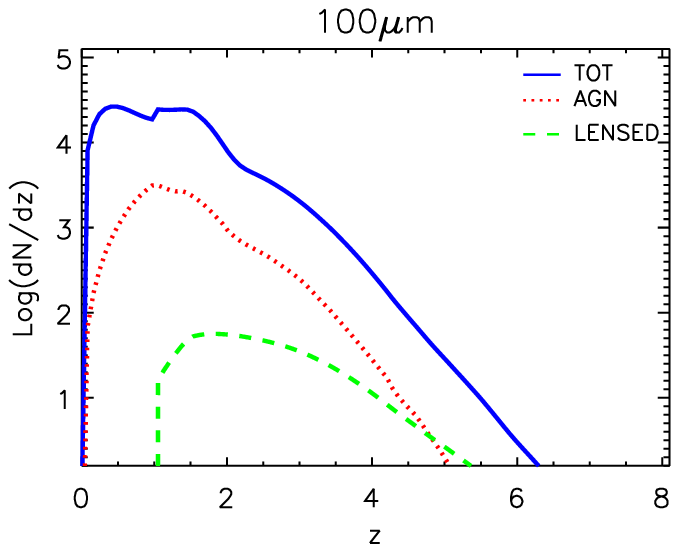}
\includegraphics[trim=3.7cm 0.65cm 1.6cm 0.25cm,clip=true,width=0.31\textwidth, angle=0]{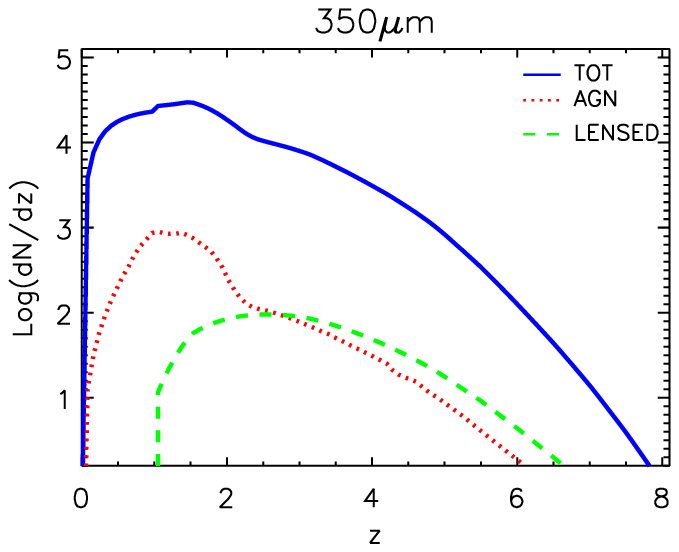}\\
\end{tabular}
\end{center}
\caption
{ \label{fig:zdistr} Predicted redshift distributions at 30, 100 and $350\,\mu$m (from left to right) at the detection limits of the reference survey (0.14, 0.21, 0.33)\,mJy, respectively. The deep hollow at $z\sim 2$ in the $30\,\mu$m distribution is due to the strong silicate absorption around $9.7\,\mu$m. The inflection point at $z\simeq 2$--2.5, seen at 100 and $350\,\mu$m, corresponds to the transition from starburst to proto-spheroid dominance.
 }
\end{figure*}

\begin{figure*}
\begin{center}
\begin{tabular}{c}
\includegraphics[trim=3.45cm 0.65cm 1.6cm 0.25cm,clip=true,width=0.32\textwidth, angle=0]{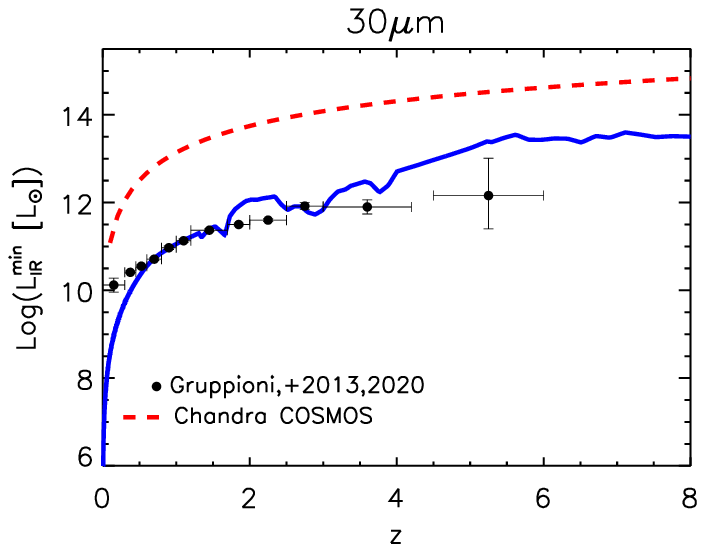}
\includegraphics[trim=3.45cm 0.65cm 1.6cm 0.25cm,clip=true,width=0.32\textwidth, angle=0]{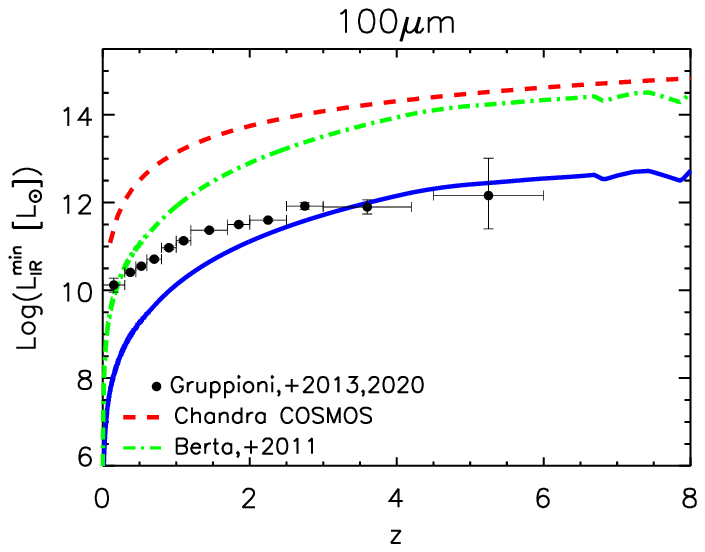}
\includegraphics[trim=3.45cm 0.65cm 1.6cm 0.25cm,clip=true,width=0.32\textwidth, angle=0]{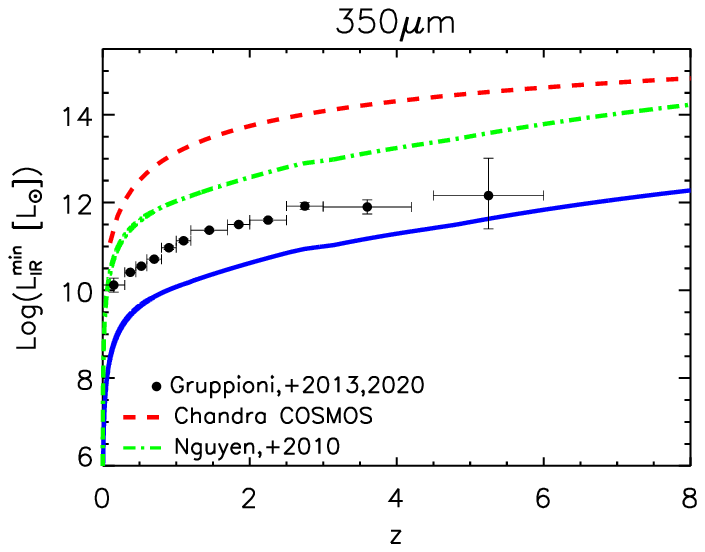}\\
\includegraphics[trim=3.45cm 0.65cm 1.6cm 0.25cm,clip=true,width=0.32\textwidth, angle=0]{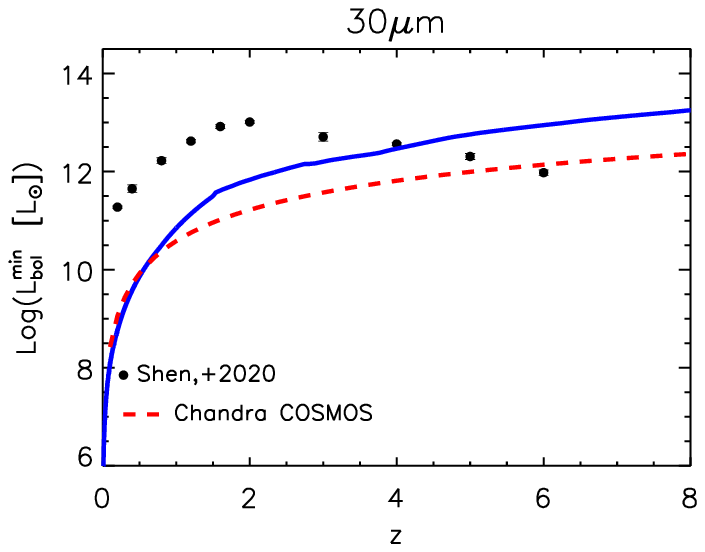}
\includegraphics[trim=3.45cm 0.65cm 1.6cm 0.25cm,clip=true,width=0.32\textwidth, angle=0]{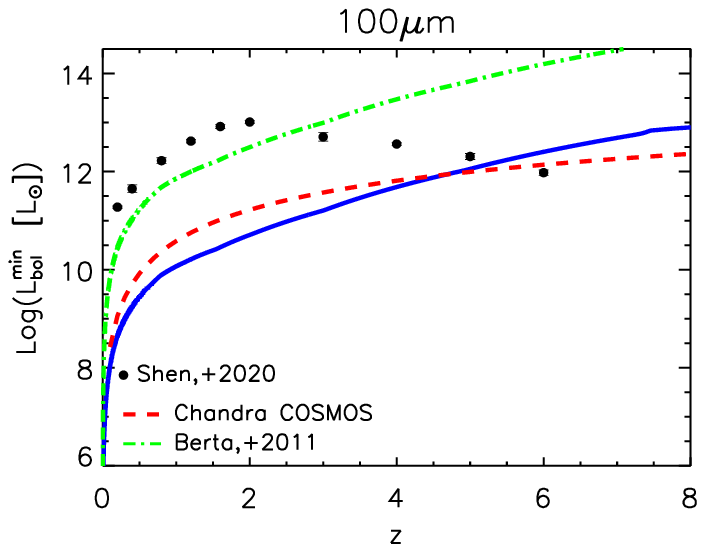}
\includegraphics[trim=3.45cm 0.65cm 1.6cm 0.25cm,clip=true,width=0.32\textwidth, angle=0]{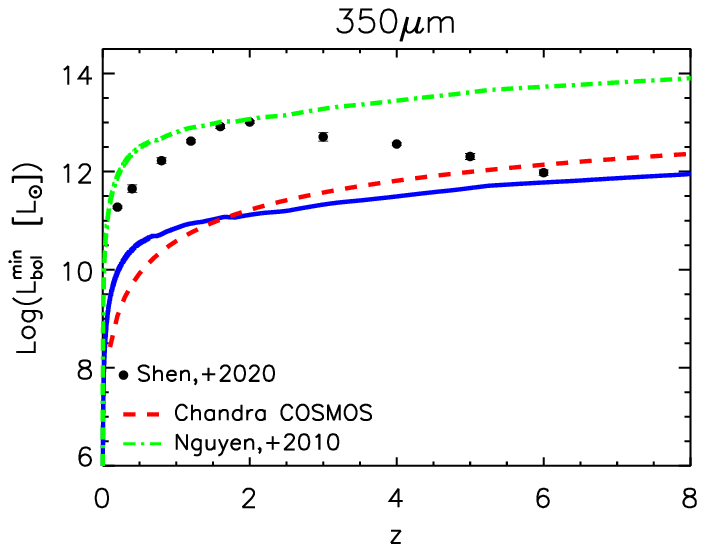}
\end{tabular}
\end{center}
\caption
{ \label{fig:Lmin} \textit{Upper panels:} Predicted minimum IR (8--$1000\,\mu$m) luminosity of star-forming galaxies detected by the reference survey at 30, 100 and $350\,\mu$m (blue curves) as a function of redshift. The dot-dashed green line shows, for comparison, the minimum luminosity corresponding to the 80\% completeness of the PEP survey of the COSMOS field \citep[9\,mJy;][]{Berta2011} at $100\,\mu$m and that corresponding to the \textit{Herschel}/SPIRE $5\,\sigma$ confusion limit at $350\,\mu$m \citep[31.5\,mJy;][]{Nguyen2010}. The points with error bars show the characteristic luminosity, $L_\star(z)$, of the Schechter fit of the IR luminosity function by \citet{Gruppioni2013}; the point at $4.5 \le z \le 6$ is from \citet{Gruppioni2020}. The dashed red line shows an estimate of the IR luminosity of the star-forming galaxy hosting an AGN having a 2--10\,keV X-ray luminosity at the 90\% completeness detection limit of the Chandra COSMOS legacy survey \citep[$7.8\times 10^{-15}\,\hbox{erg}\,\hbox{cm}^{-2}\,\hbox{s}^{-1}$; Table\,2 of][]{Puccetti2009}, as a function of $z$. To compute the K-correction we assumed a power-law energy spectrum with spectral index of $-0.4$, the slope of the cosmic X-ray background in this energy range; this represents an effective spectral index taking into account both obscured and unobscured AGN.
\textit{Lower panels:} Minimum bolometric luminosity of AGN detectable by the reference survey at the three wavelengths as a function of redshift (solid blue line). The dashed red line shows, for comparison, an estimate of the minimum bolometric luminosity of AGN detected in X-rays by the Chandra COSMOS legacy survey, as a function of $z$; it was obtained applying the bolometric correction of 22.4  \citep[from][]{Chen2013} to the 2--10\,keV luminosity computed as described above. The data points show the characteristic AGN bolometric luminosity up to $z=6$ derived by \citet{Shen2020}. Similarly to the upper panels, the dot-dashed green line shows, for comparison, the minimum bolometric luminosity corresponding to the 80\% completeness of the PEP survey of the COSMOS field \citep[9\,mJy;][]{Berta2011} at $100\,\mu$m and that corresponding to the \textit{Herschel}/SPIRE $5\,\sigma$ confusion limit at $350\,\mu$m \citep[31.5\,mJy;][]{Nguyen2010}. }
\end{figure*}

\begin{figure*}
\begin{center}
\begin{tabular}{c}
\includegraphics[trim=3.4cm 0.4cm 1.4cm 0.4cm,clip=true,width=0.45\textwidth, angle=0]{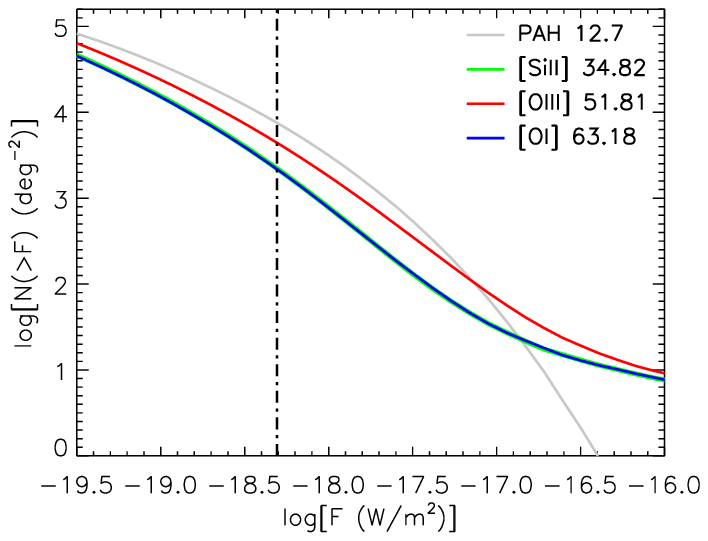}
\includegraphics[trim=3.4cm 0.4cm 1.4cm 0.4cm,clip=true,width=0.45\textwidth, angle=0]{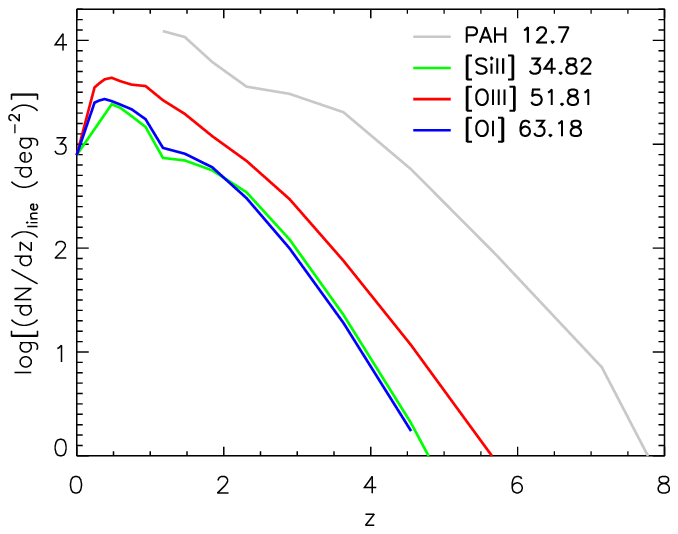}\\
\includegraphics[trim=3.4cm 0.4cm 1.4cm 0.4cm,clip=true,width=0.45\textwidth, angle=0]{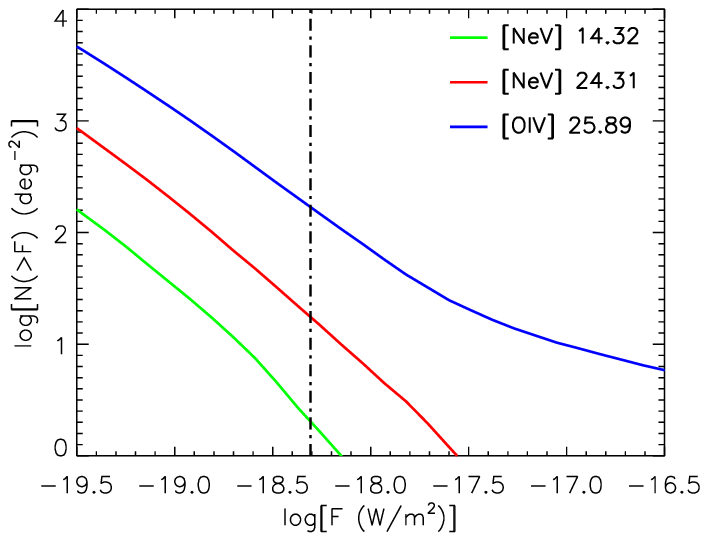}
\includegraphics[trim=3.4cm 0.4cm 1.4cm 0.4cm,clip=true,width=0.45\textwidth, angle=0]{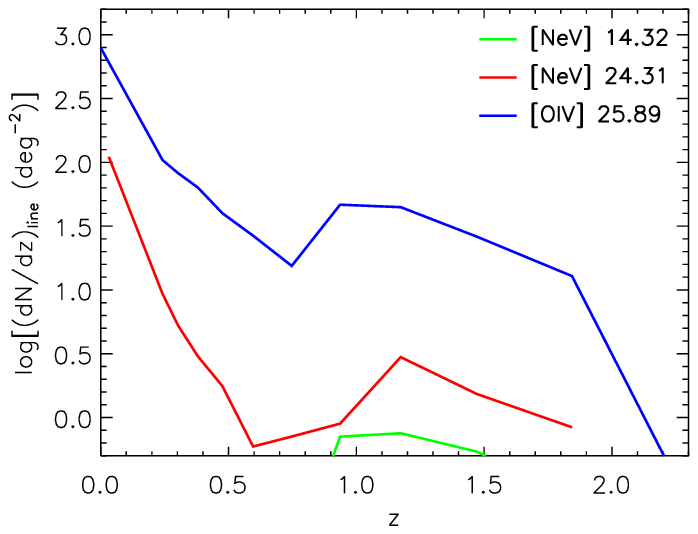}
\end{tabular}
\end{center}
\caption
{\label{fig:lines} Predicted integral number counts ({\it left}) and redshift distributions ({\it right}) for sources detected in line emission by the reference survey. In the top-left panel the integral counts of the [SiII] line are superimposed on those of the [OI] line. The dot-dashed vertical black lines in the left-hand panels show the $5\,\sigma$ detection limit (mean detection limit for the reference survey and for the fine-structure lines among the four spectral bands). \textit{Upper panels:} Star-forming galaxies detected in one or more bright lines, with spectral resolution $R=500$ for fine structure lines and $R=50$ for the PAH band. The inflection points of the redshift distributions are due to the transition from the first band (25--$50\,\mu$m) to the more sensitive second band (50-$100\,\mu$m). \textit{Lower panels:} Same as in the upper panel but for AGN lines. The transition from the first to the second band is responsible for the bump at $z\gtrsim 1$.}
\end{figure*}

Figure\,\ref{fig:counts_continuum} shows the integral counts predicted by the model ({described in the previous subsection}) at 30, 100, and $350\,\mu$m, including the contributions of AGN and of strongly lensed galaxies. The vertical black lines correspond to the $5\,\sigma$ detection limits of our reference survey.

At these limits, the counts have already substantially flattened, so going wider is far more advantageous than going deeper. The FIR wavelength range encompasses the peak of the cosmic infrared background, occurring at $\sim 200\,\mu$m \citep{Fixsen1998}. We expect that the reference survey will yield approximately 8,500, 46,000, and 61,000 $\ge 5\,\sigma$ detections at 30, 100, and $350\,\mu$m, respectively. These sources contribute $\sim$ 0.02, 0.35, and 0.75 $\hbox{MJy}\,\hbox{sr}^{-1}$ at 30, 100, and $350\,\mu$m, respectively, i.e. $\sim$ 60\%, 87\%, and 84\% of the cosmic infrared background intensity predicted by the model, which is consistent with the fit to the observational intensity estimate given by eq.~(5) of \citet{Fixsen1998}.

The number of strongly lensed galaxies is small at $30\,\mu$m ($\simeq 20$ sources, 0.23\%) and increases to $\simeq 120$ (0.26\%) at $100\,\mu$m and to $\simeq 260$ (0.42\%) at $350\,\mu$m. At the latter wavelength 120 strongly lensed galaxies are detected at $\ge 20\,\sigma$, implying that multiple images will be visible \citep[see the discussion in the Appendix of][]{Mancuso2015}.
Taking advantage of the gravitational stretching of the images it will be possible to investigate their internal structure.

At $30\,\mu$m the global AGN fraction of detected sources is $\simeq 9\%$, increasing to $\simeq 25\%$ above 3\,mJy. Although the fraction steadily decreases with increasing wavelength due to the warmer AGN MIR/FIR SED compared to that of dusty galaxies, the total AGN number peaks at around $100\,\mu$m with $\simeq 4,100$ detections ($\simeq 740$ detections at $30\,\mu$m; $\simeq 1,200$ at $350\,\mu$m). We note that these AGN counts are computed in terms of the flux density from the active nucleus only. The number of sources hosting an AGN detectable in other wavebands (X-ray, optical, near/mid-IR) is substantially larger but difficult to quantify.

Figure\,\ref{fig:zdistr} shows the predicted redshift distributions at 30, 100 and $350\,\mu$m of sources above the $5\,\sigma$ detection limits of the reference survey. The distributions extend to higher and higher redshifts with increasing wavelength, reaching the reionization epoch at $350\,\mu$m. There are $\simeq 160$, 1,100, and 7,600 galaxies at $z\ge 3$, respectively, at the three wavelengths.

Figure\,\ref{fig:Lmin} shows the minimum IR luminosity of star-forming galaxies and the minimum bolometric luminosity of AGN, detectable at $5\,\sigma$ by the reference survey. The large improvement over \textit{Herschel} surveys (dot-dashed green lines) allows a big step forward in the observational investigation of the co-evolution of the SFR and of the accretion rate onto the supermassive black hole, as discussed in the next section. The minimum detectable IR luminosity of galaxies is much lower than that corresponding to the minimum X-ray luminosity detected by the deep Chandra-COSMOS Legacy Survey \citep{civano2016}, according to the SFR/accretion rate correlation by \citet{Chen2013}. At the longest wavelengths, exemplified by the $350\,\mu$m panel, the survey reaches luminosities substantially fainter than the characteristic IR luminosity, $L_\star(z)$, of star-forming galaxies over the redshift range at which it has been estimated. This will allow us to put on much more solid ground the determination of the redshift-dependent IR luminosity function and the corresponding obscured cosmic star-formation rate density.
This will crucially complement previous studies at the high-$z$ end, which so far have been mostly based on UV-selected samples of massive galaxies (see, e.g., ALPINE at $z =4-6$, \citealt{Gruppioni2020,Khusanova2021}, and REBELS at $z\sim 7$, \citealt{Algera2023,Barrufet2023}; see also \citealt{Zavala2021}).

The lower panels of Fig.\,\ref{fig:Lmin} show that the reference survey has a depth comparable to that of the \textit{Chandra}-COSMOS Legacy Survey \citep{civano2016} in terms of AGN bolometric luminosity, making it possible to get a comprehensive view of the AGN SED. Except at the highest redshifts, both surveys reach luminosities well below the characteristic bolometric luminosity $L_{\rm bol\,AGN,\star}(z)$ derived by \citet{Shen2020} up to $z=6$.

%The predicted integral counts of galaxies and AGN detected in lines are shown in Fig.\,\ref{fig:lines}, together with the redshift distributions at the detection limit of the reference survey. \textbf{In this Figure, on the top, we display some examples of the brightest SF lines, in which we expect to detect most of the SFGs. On the bottom, we show the three brighest AGN lines}. The strong PAH band at $12.7\,\mu$m\footnote{\textbf{We expect comparable numbers of detections in the PAH lines listed in Table\,\ref{tab:lines} and we choose the PAH\,$12.7\,\mu$m as representative of all of them.}} will be detected  \textbf{from z$\sim$1(that is the lowest reachable redshift among the PAH lines)} up to $z \gtrsim 6$; the [OIII]\,$51.81\,\mu$m line up to $z \gtrsim 5$,  \textbf{[SiII]}\,$34.82\,\mu$m and [OI]\,$61.18\,\mu$m up to $z\gtrsim 4$.

{The left-hand panels of Fig.\,\ref{fig:lines} show the predicted integral counts of star-forming galaxies and AGN (upper and lower panel, respectively) detected in their brightest lines. The right-hand panels show the corresponding redshift distributions at the detection limit of the reference survey. We note that the PAH band at $12.7\,\mu$m\footnote{{We expect comparable numbers of detections in the other PAH lines listed in Table\,\ref{tab:lines} and we choose the PAH\,$12.7\,\mu$m as representative of all of them.}} is detectable only at $z\gtrsim 1$.  We expect a significant number of detections in this line up to $z>6$; in the [OIII] $51.81\,\mu$m line up to $z \gtrsim 5$; in the [SiII] $34.82\,\mu$m and in the  [OI] $61.18\,\mu$m lines, up to $z \gtrsim 4$.}

The AGN lines are relatively weak. The slope of the counts below the detection limit of the reference survey is steeper than that of star-forming galaxies, implying that in this case there is a gain in going deeper. We expect fewer than one hundred detections in the [OIV]\,$25.89\,\mu$m line at $z \lesssim 2$ with the reference survey.  As mentioned in Sect.\,\ref{sect:introduction}, dust-enshrouded AGN at higher redshift may be uncovered through the intensity ratios of PAH features. Also, information on AGN can be retrieved by stacking fluxes at the positions of sources detected in the continuum.

\begin{figure}
\begin{center}
\begin{tabular}{c}
\includegraphics[height=5.5cm]{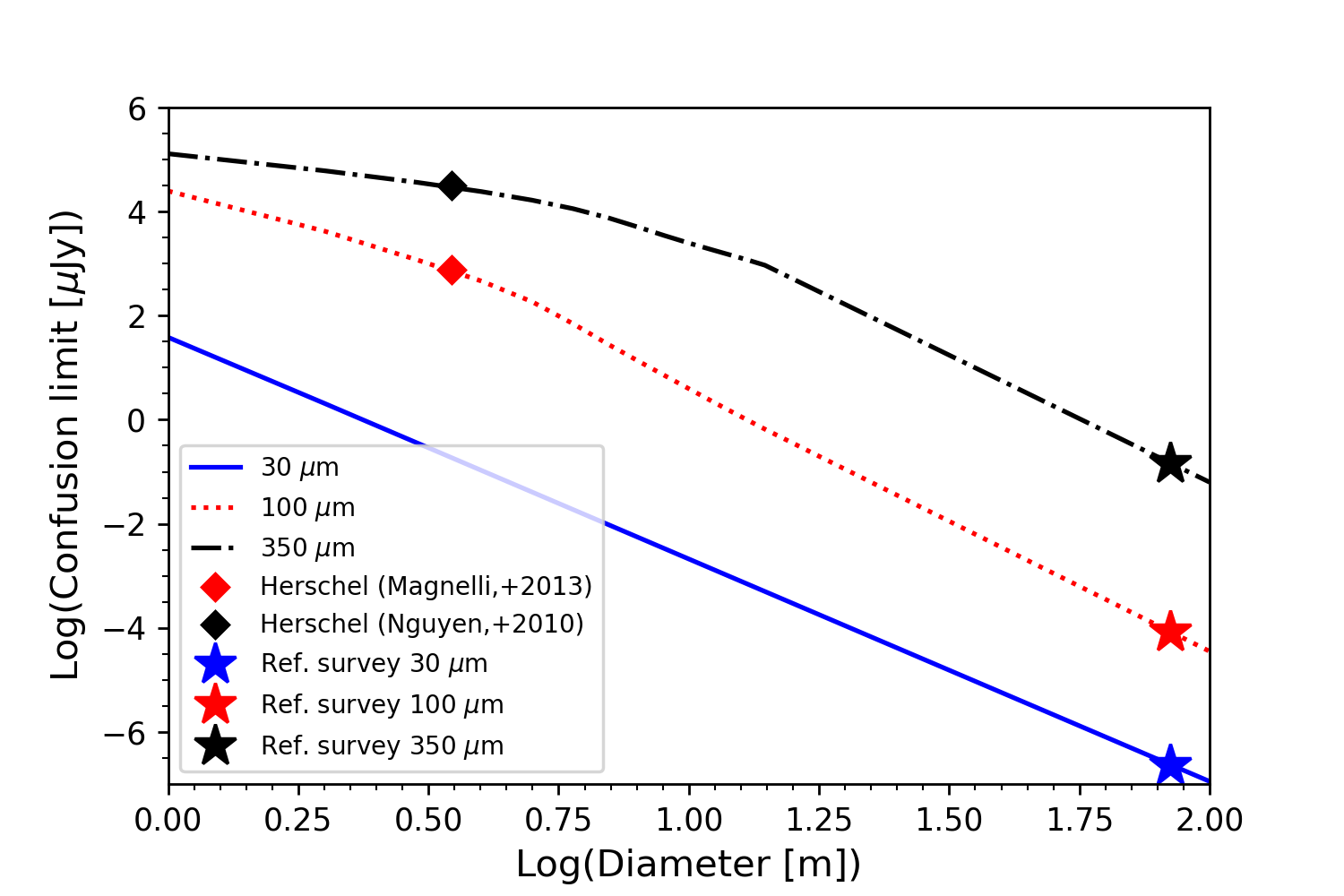}
\end{tabular}
\end{center}
\caption
{\label{fig:ConfLim}
Confusion limit, $S_{\rm conf}$ ($5\,\sigma$), at 30, 100 and $350\,\mu$m as a function of the effective telescope diameter or maximum interferometer baseline, $D$. The diamonds show the measured confusion limits of \textit{Herschel}/SPIRE at $350\,\mu$m \citep[31.5\,mJy, black diamond;][]{Nguyen2010} and of \textit{Herschel}/PACS at $100\,\mu$m \citep[0.75\,mJy, red diamond;][]{Magnelli2013}. The improved resolution of the interferometer results in a dramatic fall of confusion limits. {The stars represent the confusion limits for the proposed survey.} The model, used to compute $S_{\rm conf}$ as a function of $D$ at the three wavelengths, {shown by the solid
blue line, by the dotted red line, and by the dot-dashed black line,} yields $S_{\rm conf}= 2.3\times10^{-7}$, $8.3\times10^{-5}$, and $0.14\,\mu$Jy at 30, 100, and $350\,\mu$m, respectively. Confusion is negligible at these levels.  }
\end{figure}

\section{Discussion}\label{sect:discussion}

A sensitive FIR space mission achieving sub-arcsecond resolution is an essential complement to JWST and ALMA if we are to definitively address the Astro2020 Decadal Survey {\citep{decadalsurvey2021}} imperative to understand the co-evolution of galaxies and their central supermassive black holes. Such a mission would allow us to study the dusty interstellar medium (ISM) of star-forming galaxies and heavily obscured AGN at crucial wavelengths for individual sources (both galaxies and AGN) out to high redhsifts. Continuum FIR observations are essential to determine the total IR luminosity of star-forming galaxies.

As illustrated by Fig.\,\ref{fig:ConfLim}, the much better angular resolution of the proposed SPIRIT interferometer effectively eliminates extragalactic source confusion. Figure\,\ref{fig:ConfLim} shows the dependence on the effective telescope diameter or maximum interferometer baseline, $D$, of $S_{\rm conf}=5\,\sigma_{\rm conf}$, $\sigma_{\rm conf}$ being the rms confusion noise given by
\begin{equation}
\label{eq:conf}
\sigma^2_{\rm conf} = \omega_{\rm eff} \int_{0}^{S_{\rm conf}} \frac{dN}{dS}\,S^2\,dS\, ,
\end{equation}
where, assuming a Gaussian beam, the effective solid angle is
\begin{equation}
\label{eq:om_eff}
\omega_{\rm eff} = \frac{\pi}{2\ln 2}\left(\frac{\hbox{FWHM}}{2}\right)^2,
\end{equation}
with $\hbox{FWHM}=1.2\,\lambda/D$. The three curves in Fig.\,\ref{fig:ConfLim} show, from bottom to top, $S_{\rm conf}(D)$ for $\lambda = 30$, 100, and $350\,\mu$m, respectively.

A FIR interferometer will allow us to extend the determination of the IR luminosity functions of star-forming galaxies and of AGN down to much fainter luminosities, whereas smaller, confusion-limited telescopes see only relatively bright cosmological sources and may underestimate or overestimate line strengths. We showed that, for the considered observing time, the number of detections is maximized by a survey of $0.5\,\hbox{deg}^2$, approximately the same area as the COSMOS-Web JWST treasury program  \citep{Casey2023}, covering wavelengths from 25 to 400\,$\mu$m. We calculated the expected source detections for a survey of 1000\,hr (allowing for 90\% observing efficiency) over such an area, down to the $5\,\sigma$ detection limits of (0.10, 0.14, 0.21, 0.33)\,mJy in the continuum in the 25--50, 50-100, 100--200 and 200--400$\,\mu$m wavelength intervals, respectively.

% CANDELS \citep{Grogin2011, Koekemoer2011}

At $350\,\mu$m this reference survey will detect $\sim 61,000$ star-forming galaxies out to redshift $z\simeq 6$, and perhaps to $z=7$-8; about 7,600 galaxies will be at $z>3$. A few hundred galaxies should be strongly lensed; about half of these will be detected at $\gtrsim 20\,\sigma$ so that it will be possible to {take advantage of the gravitational stretching of the images to further improve the resolution at which the internal structure can be studied.}

A comparison, shown in Fig.\,\ref{fig:Lmin}, with the characteristic IR luminosity of star-forming galaxies as a function of redshift, $L_{\rm IR,\star}(z)$, estimated by \citet{Gruppioni2013} and  \citet{Gruppioni2020}, shows that, at least at the longest wavelengths, the proposed survey will reach luminosities fainter than $L_{\rm IR,\star}(z)$, thus allowing us to investigate typical star-forming galaxies at high $z$. This survey will also resolve the sources contributing up to $\sim$87\% of the cosmic infrared background at its peak.

In the obscured AGN case, the FIR interferometer will go much deeper than ALMA and provide unique information on the global energetics of dust-obscured nuclear activity, hence on the AGN accretion history. The number of AGN detections is highest around $100\,\mu$m, where the reference survey is predicted to detect $\sim 4,100$ such objects at redshifts of up to $z\sim4$--4.5. At longer wavelengths, we expect AGN detections up to $z\sim 5.5$--6.

Again at the longest FIR wavelengths, the AGN bolometric luminosity function can be determined down to below the characteristic bolometric luminosity $L_{\rm bol\,AGN,\star}(z)$ derived by \citet{Shen2020} up to $z=6$ (lower panels of Fig.\,\ref{fig:Lmin}). This amounts to getting a direct, solid, quantitative assessment of the cosmic star formation and obscured accretion history; presently both of them are poorly constrained for $z> 2$.

Another possible breakthrough is the understanding of complex facets of galaxy-AGN co-evolution. The tight correlations between the mass of the {super-massive black hole (SMBH)} and the global properties of the spheroidal component of the host galaxy \citep{FerrareseFord2005,KormendyHo2013}, and the broad parallelism of the cosmic history of black hole and stellar mass growth \citep{Shankar2009} imply a strong evolutionary connection. However, the physical processes responsible for it are still being debated. These include various kinds of feedback mechanisms \citep{SilkRees1998, Fabian1999, King2003, Granato2004, Murray2005, McQuillinMcLaughlin2012,Farrah2012,Fabian2012}, capable of controlling the star formation and the SMBH accretion rate. Alternatively, the evolution of both components can be regulated by galaxy properties that determine how the gas feeds star formation and accretion \citep{Angles-Alcazar2015, Ni2021}. Another possible explanation links the correlations to the effect of mergers \citep{JahnkeMaccio2011, GrahamSahu2023}.

Also, models predict different star formation and accretion histories as a function of galactic age, with feedback either inducing or quenching star formation, but observational verifications have remained elusive. A direct test of galaxy/AGN co-evolution models \citep{Granato2004, DiMatteo2005, Lapi2006, Lapi2014, Hirschmann2014, Weinberger2017} can be performed by studying the correlation between star-formation and accretion rates \citep{Chen2013, Delvecchio2015, Rodighiero2015, Lanzuisi2017, Aird2019, Suh2019, Carraro2020, MountrichasShankar2023, Lopez2023}. To this end we can exploit the X-ray luminosity, a clean tracer of nuclear emission, i.e. of the black hole accretion rate (BHAR), and the far-IR luminosity, an efficient indicator of the host galaxy SFR. Most of the cited studies have targeted the COSMOS field, taking advantage of the extensive multi-frequency photometry and spectroscopic information available with a unique combination of depth and area ($\simeq 2\,\hbox{deg}^2$).

The \textit{Chandra}-COSMOS Legacy Survey \citep{civano2016} has provided a deep, uniform X-ray coverage of the field. The field was also covered by the \textit{Herschel} Photodetector Array Camera and Spectrometer (PACS) Evolutionary Probe (PEP) survey \citep{Lutz2011, Berta2011} and by the  \textit{Herschel} Multi-tiered Extragalactic Survey \citep[HerMES;][]{Oliver2012}. However, a minor fraction of sources have both \textit{Herschel} and X-ray  measurements. For example, only 27\% of the X-ray selected sample by \citet{Suh2019} have a detection at least at one \textit{Herschel} wavelength, implying that estimates of the SFR are quite uncertain. Conversely, we have found that only 12\% of galaxies detected at $250\,\mu$m by the HerMES survey of the COSMOS field have a \textit{Chandra} detection within 10 arcsec. Thus, the study of star-formation in galaxies hosting AGN or of the AGN activity in star-forming galaxies has been so far hampered by selection biases (the different selections favour either the brighter X-ray sources or the higher SFRs that may not be representative of the general population). Various techniques (e.g., stacking, SED fitting based on templates) have been applied to make up for missing data. Conflicting results have been obtained. Some studies reported evidence of a correlation between BHAR and SFR \citep{Chen2013, Aird2019, Carraro2020, Torbaniuk2021, MountrichasShankar2023}; others concluded that SFRs are independent of AGN activity \citep{Azadi2015, Stanley2015, Suh2019, Symeonidis2022}; still others found a lack of star formation in host galaxies of X-ray quasars \citep{Barger2015}.

While the correlation between SFR and BHAR provides valuable insights, it's crucial to consider additional factors, such as stellar masses, ages, and metallicities, to draw more definitive conclusions about the underlying physical processes. Interferometric surveys can provide robust stellar masses and FIR line based ages and metallicities, simultaneously with SFRs and BHARs.

As illustrated by Fig.\,\ref{fig:Lmin}, the reference survey will reach, at all FIR wavelengths and at all redshifts, IR luminosities well below those of star-forming galaxies hosting AGN with a 2--10\,keV X-ray luminosity at the 90\% completeness detection limit of the Chandra COSMOS legacy survey, according to the correlation between AGN bolometric luminosity and SFR reported by \citet{Chen2013}. Although such correlation has a large dispersion and even its reality is controversial, the reference survey would allow decisive progress on our understanding of galaxy-{BH} co-evolution. In particular, we will be able to derive information on growth timescales of stellar and black hole masses, and on the relative importance of positive and negative feedback.

% X-ray surveys. CHANDRA COSMOS LEGACY SURVEY (Civano et al. 2016). Indice spettrale -0.4 (2-10 keV0 90\% completeness limit $7.8\times 10^{-15}\,\hbox{erg}\,\hbox{cm}^{-2}\,\hbox{s}^{-1}$. Positional uncertainty $\simeq 1''$. Bolometric correction (22.4) from Chen et al. (2013).
%
The spectroscopy we envisage will measure the physical conditions of dust-obscured atomic and ionized gas phases. The MIR/FIR fine-structure lines provide unique information on the dust-obscured phases of the ISM, which by definition are invisible in the optical. The reference survey will detect thousands of galaxies in the [OI]\,$63.18\,\mu$m, [SiII]\,$34.82\,\mu$m, and [OIII]\,$51.81\,\mu$m lines. The [OI] line is an important coolant of neutral gas; we predict about one thousand detections in this line up to $z\simeq 1$, a few hundred at $1< z < 3$ and some detections up to $z\simeq 4$. The [Si] line comes from moderately ionized gas; the redshift distribution of galaxies detected in this line is similar to that of the [OI] line. The [OIII] line traces dust in the ionized phase; it is detectable by the reference survey up to $z \simeq 5$, but its redshift distribution peaks at $z\lesssim 1$ where we expect thousands of detections. The strong PAH bands are visible up to $\simeq 6$ and are good SFR measures.
The lines with high excitation potential, such as [OIV]\,$25.89\,\mu$m and [NeV] at 14.32 and $24.31\,\mu$m, are indicative of AGN activity. The reference survey will detect the [OIV] line for almost one hundred AGN at $z\lesssim 2$. The line intensity is well correlated with the black hole accretion rate. Simultaneous measurements of this line and of SFR tracers will provide insights on galaxy-AGN co-evolution.

The spectral resolution of an interferometer like SPIRIT will allow us to exploit the brightest FIR lines to measure gas velocities, shedding light on inflows and outflows and on the interplay of different gas phases. Powerful outflows, driven by supernova explosions or by AGN activity, have a key role in controlling galaxy evolution via positive or negative feedback.

%SOME WORDS ABOUT HOW THE REFERENCE SURVEY WOULD ALLOW THE DETECTION OF A SIZEABLE SAMPLE OF FIR GALAXIES AND AGN. SOMETHING ABOUT THE FRACTION OF COSMOS X-RAY SOURCES THAT WOULD BE DETECTED.
%COMPARE TO WHAT WAS POSSIBLE WITH HERSCHEL (IT SHOULD BE MUCH BETTER).
%SEDS THAT COULD BE FITTED BECAUSE WE KNOW THE RIGHT COUNTERPARTS.
%HOW MANY OBJECTS AT Z>3.
%THE IMPACT ON SCIENCE--WHAT WE HOPE TO FIND.
%TRY TO JUSTIFY THE STATEMENT THAT WITHOUT SPICE, WE CANNOT UNDERSTAND THE CO-EVOLUTION OF BLACK HOLES AND GALAXIES.
%(SOME OF THIS MATERIAL IS ALREADY HERE, BUT LATER IN THIS SECTION)

\section{Conclusions}\label{sect:conclusions}

In summary, a FIR space mission with high angular resolution is crucial for advancing our understanding of the co-evolution of galaxies and supermassive black holes. The key findings from our analysis are summarized in the following points:
\begin{itemize}
\item An astronomical background-limited small telescope preferentially detects bright (luminous, low-$z$) galaxies.
\item A FIR high-resolution interferometer significantly reduces extragalactic source confusion and is crucial to address the co-evolution of galaxies and supermassive black holes, providing key insights that complement JWST and ALMA findings.
\item The proposed FIR survey can detect tens of thousands of star-forming galaxies and thousands of AGN up to $z \sim 6$, in multiple FIR lines (e.g., [CII], [OI], [CI]) and continuum.
\item The survey can reach luminosities below the characteristic IR luminosity at high $z$, extending IR luminosity function determinations to much fainter levels and resolving up to 87\% of the cosmic infrared background.
\item The synergy of high spectral resolution, line sensitivity, and extensive spectral coverage would enable detailed insights into the physical properties — such as temperature, density, and metallicity — of the ISM up to high redshift.
\end{itemize}

\begin{acknowledgments}

{We are grateful to the anonymous referee for a careful reading of the manuscript and many useful comments}. M.B. acknowledges support from INAF under the mini-grant ``A systematic search for ultra-bright high-z strongly lensed galaxies in Planck catalogues''. {We thank Dr. Alexander Kashlinsky for developing the original sky simulation code cited in Section\,\ref{sect:confusion} and Ms. Aláine Lee for assisting with code modifications during a summer internship at NASA’s Goddard Space Flight Center.}

\end{acknowledgments}

\bibliography{ExtragalSurv}{}
\bibliographystyle{aasjournal}

%%%%% Biographies of authors %%%%%

%\vspace{2ex}\noindent\textbf{First Author} is an assistant professor at the University of Optical Engineering. He received his BS and MS degrees in physics from the University of Optics in 1985 and 1987, respectively, and his PhD degree in optics from the Institute of Technology in 1991.  He is the author of more than 50 journal papers and has written three book chapters. His current research interests include optical interconnects, holography, and optoelectronic systems. He is a member of SPIE.

%\vspace{1ex}
%\noindent Biographies and photographs of the other authors are not available.

%\end{spacing}
\end{document}